\documentclass[preprint]{aastex}
\begin{document}

\title{Kinematics and Physical Conditions of the Innermost Envelope in B335}

\author{Hsi-Wei Yen\altaffilmark{1,2}, Shigehisa Takakuwa\altaffilmark{2}, and Nagayoshi Ohashi\altaffilmark{2}}
\altaffiltext{1}{Institute of Astrophysics, National Taiwan University, Taipei 10617, Taiwan}
\altaffiltext{2}{Academia Sinica Institute of Astronomy and Astrophysics, P.O. Box 23-141, Taipei 10617, Taiwan} 

\begin{abstract}
We made C$^{18}$O (2--1) and CS (7--6) images of the protostellar envelope around B335 with a high spatial dynamic range from $\sim$10000 to $\sim$400 AU, 
by combining the Submillimeter Array and single-dish data. 
The C$^{18}$O emission shows an extended ($\sim$10000 AU) structure as well as a compact ($\sim$1500 AU) component concentrated at the protostellar position. 
The CS emission shows a compact ($\sim$900 AU) component surrounding the protostar, 
plus a halo-like ($\sim$3000 AU) structure elongated along the east-west direction. 
At higher velocities ($|\Delta V|$ $\gtrsim$0.3 km s$^{-1}$), 
the CS emission is stronger and more extended than the C$^{18}$O emission. 
Physical conditions of the envelope were examined through an LVG model. 
At $|\Delta V|$ $\gtrsim$0.3 km s$^{-1}$, 
the gas temperature is higher ($>$40 K) than that at $|\Delta V|$ $\lesssim$0.3 km s$^{-1}$, whereas the gas density is lower ($<$10$^{6}$ cm$^{-3}$).  
We consider that the higher-temperature and lower-density gas at $|\Delta V|$ $\gtrsim$0.3 km s$^{-1}$ is related to the associated outflow,  
while the lower-temperature and higher-density gas at $|\Delta V|$ $\lesssim$0.3 km s$^{-1}$ is the envelope component. 
From the inspection of the positional offsets in the velocity channel maps, 
the radial profile of the specific angular momentum of the envelope rotation in B335 was revealed at radii from $\sim$10$^{4}$ down to $\sim$10$^{2}$ AU. 
The specific angular momentum decreases down to the radius of $\sim$370 AU, 
and then appears to be conserved within that radius. 
A possible scenario of the evolution of envelope rotation is discussed.

\end{abstract}


\section{Introduction}
Dense-gas ($\geq$10$^{4-5}$ cm$^{-3}$) condensations in dark molecular clouds are sites of low-mass star formation \citep{And00, Mye00}. 
Previous millimeter interferometric observations have revealed rotating and infalling gas motions in dense-gas condensations associated with known protostellar sources, 
so-called ``protostellar envelopes'' \citep{Oha96, Oha97a, Oha97b, Mom98}.
It is still less clear, 
however, 
how the infalling motion is terminated and the angular momentum of the envelope rotation is transferred from large ($\sim$10000 AU) to small ($\sim$a few hundred AU) radii in the envelopes (e.g., Goodman et al. 1993; Ohashi et al. 1997b). 
Therefore, it is not well understood how centrifugally supported disks with radii of a few hundred AU, often observed around T Tauri stars (e.g., Guilloteau et al. 1999; Guilloteau \& Dutrey 1998; Qi et al. 2003), are formed around protostars in those envelopes.

In order to approach these questions, we have performed Submillimeter Array (SMA)$\footnotemark$ observations of B335 (J{\o}rgensen et al. 2007; Yen et al. 2010, hereafter Paper I), 
an isolated Bok globule associated with an embedded Class 0 protostar with a bolometric luminosity of 1.5 $L_{\sun}$ (IRAS 19347+0727; Keene et al. 1980, 1983; Stutz et al. 2008). 
B335 is associated with an east-west elongated, conical-shaped molecular outflow with a size of $\sim$0.2 pc, an opening angle of $\sim$45$\degr$, and an inclination angle of $\sim$10$\degr$ from the plane of the sky \citep{Hir88}.
Along the outflow axis there are also high-velocity ($\sim$160 km s$^{-1}$) compact ($\sim$1500 $\times$ 900 AU) molecular jets seen in the $^{12}$CO (2--1) emission (Paper I), 
as well as HH objects (HH 119 A-F; Reipurth et al. 1992; G{\aa}lfalk \& Olofsson 2007).
Millimeter interferometric observations of the envelope around B335 in the H$^{13}$CO$^{+}$ (1--0) and C$^{18}$O (1--0) emission lines at an angular resolution of $\sim$6$\arcsec$ have revealed the presence of the infalling gas motion on a 3000 AU scale \citep{Cha93,Sai99}.
The envelope around B335 also exhibits a slow rotational motion at radii of $\sim$20000 AU \citep{Fre87, Sai99} and $\sim$1000 AU \citep{Sai99}.
Millimeter continuum studies of the envelope around B335 show an $r^{-1.5}$ density profile between $r$ = 60 and 3900 AU and an $r^{-2}$ density profile at $r$ $>$3900 AU \citep{Har03}, 
which is consistent with the inside-out collapse model \citep{Shu77}.
These results indicate that B335 is a prototypical low-mass protostellar source suitable for detailed studies.

\footnotetext{The Submillimeter Array (SMA) is a joint project between the Smithsonian Astrophysical Observatory and
the Academia Sinica Institute of Astronomy and Astrophysics and is funded by the Smithsonian Institute and the
Academia Sinica.}

Our SMA observation of B335 in the C$^{18}$O (2--1) emission found a compact ($\sim$1500 AU) envelope around the protostar, 
showing an infalling motion (Paper I).
The infalling velocity was measured to be $\sim$0.44 km s$^{-1}$ at a radius of 370 AU, 
and the central stellar mass was estimated to be 0.04 M$_{\sun}$.
On the other hand, there is no signature of the envelope rotation in the C$^{18}$O emission on a few hundred AU scale, 
and the upper limits of the rotational velocity and specific angular momentum of the envelope rotation were estimated to be 0.04 km s$^{-1}$ and 7 $\times$ 10$^{-5}$ km s$^{-1}$ pc at a radius of 370 AU, respectively. 
The structure and kinematics in the innermost region of the protostellar envelope, 
however, 
are not well understood with our previous millimeter molecular-line observation, 
because millimeter molecular-line observations, 
such as those in the C$^{18}$O (2--1) line, 
suffer from contamination from the surrounding low-density and low-temperature gas ($n$ $\sim$10$^{4-5}$ cm$^{-3}$ \& $T$ $\sim$10 K). 
In order to study the innermost envelope, it is required to observe submillimeter molecular-line emission, such as CS (7--6) \citep{Tak04, Tak07, Tak11}. 
In addition, 
physical conditions, i.e., temperature, density, optical depth, and excitation conditions, of the inner envelope are also uncertain in our previous observation,  
because it is impossible to derive physical conditions using only one molecular line; 
we need to observe at least two molecular lines tracing different physical conditions. 
It is important to note, however, that multi-line interferometric observations are not sufficient to derive physical conditions of the inner envelope 
because interferometric observations suffer from the effect of the missing flux. 
In order to derive physical conditions of the inner envelope at a high angular resolution, 
we need to combine interferometric and single-dish observations (e.g., Wilner \& Welch 1994; Takakuwa et al. 2007; Santiago-Garc{\'{\i}}a et al. 2009). 

In this paper, 
we report millimeter C$^{18}$O (2--1) and submillimeter CS (7--6) line observations of B335 with the SMA, Submillimeter Telescope (SMT), and Atacama Submillimeter Telescope Experiment (ASTE)$\footnotemark$. 
The SMA and single-dish data were combined to recover missing fluxes. 
From the comparison between the combined millimeter C$^{18}$O (2--1) and submillimeter CS (7--6) data,
we derived physical conditions of the envelope and the outflow in B335 from $\sim$3600 to $\sim$700 AU scales at an angular resolution of $4\farcs$8.
We also found a possible signature of the envelope rotation in B335 on $\sim$100 AU scale,
and studied radial dependence of the specific angular momentum of the envelope rotation in B335. 
Although the distance to B335 was recently newly estimated to be 90--120 pc \citep{Olo09}, 
in this paper we adopted a distance of 150 pc \citep{Stu08}, 
which was also adopted in Paper I. 
The $\sim$50\% uncertainty in the distance yields $\sim$50\% uncertainty in the estimated rotation radius and specific angular momentum discussed in the paper, 
and it does not change our main discussion on the specific angular momentum distribution and the evolutionary stage of B335.

\footnotetext{The ASTE project is driven by Nobeyama Radio Observatory, a branch of National Astronomical Observatory of Japan, in collaboration with University of Chile, and Japanese institutes including University of Tokyo, Nagoya University, Osaka Prefecture University, Ibaraki University, and Hokkaido University. Observations with ASTE were in part carried out remotely from Japan by using NTT's GEMnet2 and its partnet R$\&$E (Research and Education) networks, which are based on AccessNova collaboration of University of Chile, NTT Laboratories, and NAOJ.}

\section{Observations}
SMA observations of B335 at 230 and 342 GHz were made with seven antennas on 2005 June 24 and 14, respectively, as a part of a large SMA project (PROSAC: J\o rgensen et al. 2007).
Details of the SMA were described by \cite{Ho04}.
In the 230 GHz observation,
$^{12}$CO (2--1), $^{13}$CO (2--1), C$^{18}$O (2--1), and 1.3 mm continuum emission were observed simultaneously,
and details of the observation were described in Paper I. 
On the other hand,
H$_{2}$CO (5$_{1,5}$--4$_{1,4}$), CS (7--6), and 342 GHz continuum emission were observed simultaneously in the 342 GHz observation. 
Observational parameters of the SMA observations are summarized in Table \ref{ob}.
Our 230 and 342 GHz observations were insensitive to structures more extended than $\sim$4500 and $\sim$2000 AU at the 10\% level \citep{Wil94}, respectively. 
The MIR software package was used to calibrate the data. 
The calibrated visibility data were Fourier-transformed and CLEANed with MIRIAD \citep{Sau95} to produce images. 
In this paper, we will present the results of the C$^{18}$O (2--1; 219.56036 GHz) and CS (7--6; 342.88295 GHz) lines.

Single-dish observations of B335 in the C$^{18}$O (2-1) emission were made with SMT on 2008 November 20 and 26, 
while those in the CS (7-6) emission by using ASTE were made on 2005 August 16--20.
In the SMT C$^{18}$O observations, 
we made a 7 $\times$ 9 points mapping centered on $\alpha$(J2000) = 19$^{h}$37$^{m}$00$\fs$90, $\delta$(J2000) = 7\arcdeg34\arcmin08$\farcs$8 with a grid spacing of 15\arcsec, 
which provides a 90$\arcsec$ $\times$ 120$\arcsec$ map at Nyquist sampling. 
Each position in the map was observed at least twice, and each integration time was 60 seconds in total (on and off position). 
The central position was adopted to check the relative flux calibration, 
and the flux uncertainty was estimated to be $\sim$10\% -- 20\%. 
The telescope pointing was checked every one and half hour by observing Jupiter in the 219 GHz continuum emission. 
In the ASTE CS observations, 
the mapping center was the same as that of the SMT observations, 
and we observed 21 points at a grid spacing of 10$\arcsec$, 
providing a Nyquist-sampled map in the 40$\arcsec$ $\times$ 40$\arcsec$ region. 
Details of the ASTE observations were described by \cite{Tak07}.
Observational parameters of our single-dish observations are summarized in Table \ref{ob}.
The conversion factors from $T^{*}_{A}$ (K) to $S$ (Jy Beam$^{-1}$) of the SMT C$^{18}$O and ASTE CS observations were derived to be 66.3 and 74.8, respectively, as 
\begin{equation}
S = \frac{2k_{B}\Omega_{beam}}{\lambda^{2}}\frac{T^{*}_{A}}{\eta_{mb}},
\end{equation}
where $k_{B}$ is the Boltzmann constant, $\Omega_{beam}$ is the solid angle of the single-dish beams, $\lambda$ is the wavelength, and $\eta_{mb}$ is the main beam efficiency (68\% for SMT and 60\% for ASTE). 

We combined these single-dish data with our SMA data. 
The spatial resolutions and the noise levels in the combined images are shown in Table \ref{ob}.
Details of the combining process are described in Appendix A. 
We also made simulations of the combing process to test the feasibility and limitation of this technique. 
On the assumption of an $r^{-1}$ power-law intensity distribution, 
we found that amplitudes of the visibilities made from the single-dish images are systematically lower than the correct values due to the finite spatial sampling in the single-dish observations. 
The distortion in the combined image caused by the amplitude mismatch between the single-dish and interferometric data is, however, at most 1.5$\sigma$ of the observational noise in the outer ($r$ $>$5\arcsec) region. 
The suppression of the flux in the inner ($r$ $<$5\arcsec) region caused by the lower single-dish flux is at most 20\% of the peak flux, 
comparable to the uncertainty of the flux calibration in our single-dish and interferometric observations. 
For the case of the more complicated intensity distributions of the envelope, such as clumpy structures, 
our previous imaging simulation \citep{Tak08} shows the real clumpy structures at higher than 4$\sigma$ noise level can be still recovered after combining single-dish and interferometric data. 
Based on these results, 
we consider that the combining process is unlikely to distort the original image significantly beyond the observational uncertainties. 
Details of our simulations of the combining process are described in Appendix A.

\section{Results}
The SMA images of B335 in the C$^{18}$O (2--1) and CS (7--6) emission lines were first shown in the PROSAC paper \citep{Jor07}, 
and detailed results of the SMA C$^{18}$O (2--1) data have been presented in Paper I. 
The ASTE data of the CS (7--6) emission were first published by \cite{Tak07}. 
In this paper, 
we present and discuss the results of the combined SMA + single-dish data in the C$^{18}$O (2--1) and CS (7--6) emission lines.  
Hereafter, we adopt the peak position of the 1.3 mm continuum emission in the SMA data, 
$\alpha$(J2000) = 19$^{h}$37$^{m}$0$\fs$93, $\delta$(J2000) = 7\arcdeg34\arcmin09$\farcs$8, 
as the position of the central protostellar source (Paper I). 

\subsection{Single-dish, SMA, and Combined Images and Spectra of Millimeter C$^{18}$O and Submillimeter CS Emission Lines}
Figure \ref{c18omom0} shows three moment 0 maps of the C$^{18}$O (2--1) emission; one made by using only the SMT data (left), one made by using only the SMA data (middle), and one made by using both the SMT and SMA data (right).  
In the SMT map, the C$^{18}$O (2--1) emission was detected at a level higher
than 14$\sigma$ ($\int T_{\rm MB}dv$ = 8.4 K km s$^{-1}$) over the entire observing area of 13500 AU $\times$ 18000 AU.
By contrast, 
the SMA map shows a compact C$^{18}$O blob with a size of $\sim$1500 AU toward the protostar.
In the combined map, the C$^{18}$O emission shows a central compact component surrounded by an extended component elongated along the east-west direction.
Note that the extended emission seen in the SMT map is significantly suppressed in the combined map because of the response of the SMA primary beam.

Figure \ref{csmom0} shows CS (7--6) moment 0 maps made by using only the ASTE data (left), only the SMA data (middle), and both the ASTE and SMA data (right).
In the ASTE map, 
the CS emission was detected only at the central 3 $\times$ 3 observing points \citep{Tak07},  
suggesting that the CS (7--6) emission is much more compact than the C$^{18}$O emission. 
The different emission extents are not because of the insufficient sensitivity of the ASTE CS (7--6) observations since the noise level per unit velocity width in the ASTE data ($\sim$1.3 K km s$^{-1}$) is lower than that in the SMT data ($\sim$3.7 K km s$^{-1}$).  
Although the CS (7--6) emission was not spatially resolved in the ASTE map at an angular resolution of 22\arcsec, 
it was resolved into two peaks in the SMA map. 
One emission component is coincident with the protostar, 
and the other is at $\sim$5$\arcsec$ southwest of the protostar.
In the combined map, 
the two emission peaks found in the SMA observation appear to be surrounded by a weaker extended component with a size of $\sim$3000 AU. 
The extended component shows east-west elongation as in the case of the C$^{18}$O emission,
while the inner region ($r$ $<$900 AU) shows a central compact component plus south-east and south-west protrusions. 

From the comparisons of the single-dish and SMA line profiles, 
the missing fluxes of the SMA observations were estimated to be $\sim$80\% and $\sim$85\% for the C$^{18}$O (2--1) and CS (7--6) observations, respectively (see Figure \ref{line}). 
After the single-dish and SMA data were combined, the missing fluxes were recovered properly as shown in the combined spectra in Figure \ref{line}. 
By fitting a Gaussian function to the C$^{18}$O line profile taken with SMT, 
the centroid velocity was measured to be $V_{\rm LSR}$ = 8.34 km s$^{-1}$ (vertical dashed line in Figure \ref{line}). 
Hereafter, this velocity is adopted as the systemic velocity, 
and all the velocities in the rest of this paper are shown as relative velocities with respect to this systemic velocity ($\equiv$ $V$).    

\subsection{Velocity Structure of the C$^{18}$O (2--1) Emission}
Figure \ref{smtchannel} shows the velocity channel maps of the SMT C$^{18}$O data. 
At higher velocities ($V$ = -0.30 \& 0.28 km s$^{-1}$), 
the C$^{18}$O emission mainly arises from the central region showing elongation along the east-west direction. 
At blueshifted velocities ($V$ = -0.23 and -0.17 km s$^{-1}$), 
the peak of the C$^{18}$O emission is located mostly toward the south-east of the protostar,  
while toward the north-west of the protostar at redshifted velocities ($V$ = 0.15 and 0.22 km s$^{-1}$), 
suggesting that there is a velocity gradient from the south-east to the north-west. 

In the velocity channel maps of the combined SMT + SMA C$^{18}$O data (Fig. \ref{c18ochannel}),
the extended component with a size of $\sim$9000 AU seen in the combined moment 0 map (Figure \ref{c18omom0}) appears around the systemic velocity ($V$ = -0.08 and 0.20 km s$^{-1}$). 
The velocity structure seen in the SMT velocity channel maps cannot be identified in the combined data 
since the velocity resolution of the combined data (0.28 km s$^{-1}$) is more than four times worse than that of the SMT data (0.06 km s$^{-1}$).  
On the other hand, the central compact component appears both at the systemic and higher velocities, 
and its peak is shifted from the east to the west of the protostar as the velocity changes from blueshifted to redshifted (Paper I). 

Figure \ref{c18opv} shows the SMT and combined SMT + SMA position--velocity ($P$--$V$) diagrams of the C$^{18}$O emission in B335 along and across the outflow axis passing through the central protostellar position.
The SMT $P$--$V$ diagrams shown in green contours in the top panels of Figure 6 trace the kinematics of the extended component with a size of $\sim$9000 AU,
while the combined $P$--$V$ diagrams shown in black contours focus on the kinematics of the central compact component with a size of $\sim$1500 AU. 
The velocity gradient seen in the SMT C$^{18}$O channel maps (Fig. \ref{smtchannel}) in the south-east to north-west direction can be also identified in the the SMT $P$--$V$ diagrams as an east-west velocity gradient along the outflow axis (green contours in the top-left panel of Figure \ref{c18opv}) and a north-south velocity gradient across the outflow axis (green contours in the top-right panel of Figure \ref{c18opv}). 
The trend of the velocity gradient in the east-west direction along the outflow axis is the same as that of the associated molecular outflow, 
whereas the amount of the velocity gradient in the C$^{18}$O emission ($\sim$4.1 $\times$ 10$^{-6}$ km s$^{-1}$ AU$^{-1}$) is much smaller than that of the outflow (1.2 $\times$ 10$^{-3}$ km s$^{-1}$ AU$^{-1}$; Paper I). 
Hence, 
the east-west velocity gradient seen in the SMT C$^{18}$O data most likely reflects the envelope gas motion with a possible influence from the associated outflow. 
On the other hand, 
the velocity gradient in the north-south direction seen in the SMT $P$--$V$ diagram across the outflow axis (green contours in the top-right panel of Figure \ref{c18opv}) is not as clear as that seen in the SMT $P$--$V$ diagram along the outflow axis. 
To clarify this velocity gradient, 
we performed Gaussian fitting to the SMT C$^{18}$O spectra along the north-south direction, 
and measured the centroid velocities. 
Figure \ref{SMTspec} shows the results of the Gaussian fitting. 
It is clear that there are systematic shifts of the centroid velocities between the northern (redshifted) and southern (blueshifted) spectra. 
The amount of the north-south velocity gradient ($\sim$3.9 $\times$ 10$^{-6}$ km s$^{-1}$ AU$^{-1}$) is comparable to that on $\sim$20000 AU scale seen in the C$^{18}$O (1--0) emission ($\sim$2.8 $\times$ 10$^{-6}$ km s$^{-1}$ AU$^{-1}$; Saito et al. 1999), 
and we consider that this velocity gradient seen in the SMT data likely reflects the envelope rotation. 
These velocity gradients seen in the SMT data are not clear in the combined $P$--$V$ diagrams as seen in a comparison of the $P$--$V$ diagrams between the SMT and combined data in the top panels of Figure \ref{c18opv}, 
because the velocity resolution in the combined data ($\sim$0.28 km s$^{-1}$) is a factor of four worse than that of the SMT data ($\sim$0.06 km s$^{-1}$).

In contrast to the extended component with a narrow line width ($\sim$0.5 km s$^{-1}$), 
the central compact component has a much broader line width ($\sim$1.5 km s$^{-1}$), 
as demonstrated in the combined $P$--$V$ diagrams.
The velocity gradient from east to west in the central compact component seen in Figure \ref{c18ochannel} is clearly shown in the combined $P$--$V$ diagram along the outflow axis (bottom-left panel of Figure \ref{c18opv}). 
Although the trend of the velocity gradient in the central compact component is the same as that of the east-west velocity gradient seen on thousands of AU scale in the SMT $P$--$V$ diagram along the outflow axis, 
the amount of the velocity gradient ($\sim$3.7 $\times$ 10$^{-3}$ km s$^{-1}$ AU$^{-1}$) is three order of magnitude larger than that seen in the SMT $P$--$V$ diagram. 
This velocity gradient was interpreted as an infalling motion (Paper I).
On the other hand,
the combined $P$--$V$ diagram across the outflow axis (bottom-right panel of Figure \ref{c18opv}) does not show any clear velocity gradient on a size scale of $\sim$1500 AU.  

\subsection{Velocity Structure of the CS (7--6) Emission}
Figure \ref{cschannel} shows velocity channel maps of the ASTE + SMA CS data.
At higher velocities ($V$ = -0.96 to -0.60 and 0.46 to 0.82 km s$^{-1}$),
the east-west elongated component is seen,
while at lower velocities ($V$ = -0.42 to 0.46 km s$^{-1}$) the central compact component associated with the protostar is evident.
At $V$ = -0.42 and -0.25 km s$^{-1}$,
a secondary emission peak at the south-west of the protostar is also seen,
as already found in the CS (5--4) observation \citep{Wil00}. 
The peak of the central compact component is slightly ($\sim$1\arcsec) shifted from south to north of the protostar
as $V$ changes from -0.25 to 0.46 km s$^{-1}$, 
suggestive of a possible velocity gradient along the north-south direction. 
The amount of the peak shift is smaller than the beam size, but twice larger than the relative positional accuracy $\sim$0\farcs4 estimated from $\frac{\theta}{S/N}$ where $\theta$ is the beam size and $S/N$ is the signal-to-noise ratio.
The amount of the velocity gradient was estimated to be $\sim$2 $\times$ 10$^{-3}$ km s$^{-1}$ AU$^{-1}$ from the peak shift in the velocity channel maps. 
A hint of this possible velocity gradient in the north-south direction may be also seen in the $P$--$V$ diagram of the combined ASTE+SMA CS data across the outflow axis (right panel of Figure \ref{cspv}).  
Although the trend of the velocity gradient across the outflow axis in the central compact CS component is the same as that of the SMT C$^{18}$O emission, 
i.e., the northern part is redshifted and the southern part is blueshifted, 
the amount of the velocity gradient is larger in the CS emission ($\sim$2 $\times$ 10$^{-3}$ km s$^{-1}$ AU$^{-1}$) than that of the extended C$^{18}$O emission ($\sim$3.9 $\times$ 10$^{-6}$ km s$^{-1}$ AU$^{-1}$). 

On the other hand, 
in the $P$--$V$ diagram along the outflow axis (left panel of Figure \ref{cspv}), 
the redshifted emission mainly arises from the eastern part of the protostar, 
while the blueshifted emission arises from both the eastern and western parts, 
suggesting presence of the east (redshifted) to west (blueshifted) velocity gradient. 
The trend of this velocity gradient in the submillimeter CS (7--6) emission is opposite to that of the associated outflow and the infalling motion. 
In fact, \cite{Tak07} and \cite{Tak11} have already found such opposite velocity gradients of the submillimeter molecular lines in a few low-mass protostellar envelopes. 
They interpreted that the opposite velocity gradients along the outflow direction reflect the dispersing gas motion at the surface of the envelope perpendicular to the outflow.

In order to compare the distributions and kinematics of the millimeter C$^{18}$O (2--1) and submillimeter CS (7--6) emission, 
the CS image cube was resampled along the velocity axis to match the velocity channels with those of the C$^{18}$O image cube, 
and then the image cubes were integrated into four different velocity ranges; 
the high-velocity blueshifted ($V$ = -1.05 to -0.50 km s$^{-1}$), low-velocity blueshifted ($V$ = -0.50 to -0.06 km s$^{-1}$), low-velocity redshifted ($V$ = -0.06 to 0.61 km s$^{-1}$), and high-velocity redshifted ($V$ = 0.61 to 1.17 km s$^{-1}$) ranges.  
The images are shown in Figure \ref{kinematics}. 
In the high-velocity ranges, 
the CS emission is stronger and more extended than the C$^{18}$O emission.
In contrast, 
in the low-velocity ranges, 
the C$^{18}$O emission becomes stronger and more extended compared to the CS emission,  
and both the C$^{18}$O and CS emissions clearly show a central compact component surrounding the protostar. 
The differences between the millimeter C$^{18}$O and submillimeter CS intensities at different velocities are also demonstrated in the line profiles shown in Figure \ref{line}.
The blueshifted and redshifted peaks of the central compact C$^{18}$O component are located at the east and west of the protostar, 
respectively, 
and show a velocity gradient along the outflow axis.
On the other hand, 
the peaks of the central compact CS component show a slight ($\sim$1\arcsec) shift from south to north of the protostar as the velocity changes from blueshifted to redshifted, 
as seen in Figure \ref{cschannel}.

\section{LVG Analyses}
As shown in Figure \ref{kinematics}, 
the submillimeter CS (7--6) emission is stronger and more extended than the millimeter C$^{18}$O (2--1) emission at the high velocities ($V$ $>$0.6 \& $<$-0.5 km s$^{-1}$), 
while the C$^{18}$O emission is stronger and more extended than the CS emission at the low velocities ($V$ = 0.6 -- -0.5 km s$^{-1}$). 
The high-velocity molecular gas where the submillimeter CS emission is more dominant and the low-velocity gas where the millimeter C$^{18}$O emission is more dominant should reflect the different physical conditions. 
In order to derive their physical conditions and to discuss the origin,
we made statistical equilibrium calculations based on the Large Velocity Gradient (LVG) model \citep{Gol74, Sur77}. 
Here, 
the velocity gradient and the abundances of C$^{18}$O and CS were adopted to be 100.0 km s$^{-1}$ pc$^{-1}$, 2.5 $\times$ 10$^{-8}$, and 6 $\times$ 10$^{-10}$, respectively, 
and the justification of these values is described in Appendix B.
The beam filling factors of the both emission lines were assumed to be unity.  

To directly compare the C$^{18}$O and CS intensities and derive the physical conditions from the line intensities by the LVG calculations, 
the resampled CS image cube was convolved with the combined SMT + SMA beam of the C$^{18}$O data. 
Then, 
both the C$^{18}$O and CS image cubes were integrated over the high-velocity ($V$ = -1.05 to -0.22 \& 0.34 to 0.89 km s$^{-1}$) and low-velocity
($V$ = -0.22 to 0.34 km s$^{-1}$) ranges, 
and the image pixels were binned to have a pixel size of 4\farcs8, 
comparable to the beam size. 
Note that the integrated velocity ranges are different from those adopted in Figure \ref{kinematics}. 
In the present analysis, 
we separate the velocity components based on the velocity range of the extended and compact C$^{18}$O components (see Figure \ref{c18ochannel}).

Figure \ref{lvg} compares the C$^{18}$O and CS brightness temperatures from the resampled image cubes at the high- (red points) and low-velocity (blue points) ranges.
Only the data points within the SMA primary beam at 342 GHz are plotted. 
For the data points without the detectable C$^{18}$O or CS emission ($<$2$\sigma$), 
their 2$\sigma$ upper limits are plotted (data points with crosses). 
Most of the data points at the high velocity show that the more dominant CS emission without the detectable C$^{18}$O emission, 
and as a result, 
they are located at the left-hand side in the diagram. 
On the other hand, 
the data points at the low velocity show that the C$^{18}$O emission is more dominant, 
and as a result, 
they are located at the right-hand side with respect to the high-velocity data points. 
Solid and dashed curves show iso-thermal and iso-density contours calculated by our LVG model, respectively. 
It is clear that the high-velocity gas with the stronger CS emission shows a high temperature ($>$40 K) but a low density ($<$10$^{6}$ cm$^{-3}$), 
while the low-velocity gas with stronger C$^{18}$O emission shows a low temperature ($<$40 K) but a high density ($>$10$^{6}$ cm$^{-3}$). 
Based on the LVG results shown in Figure \ref{lvg},
we made temperature and density maps overlaid with the integrated CS emission (Fig. \ref{tnmap}). 
In addition to the difference of the temperature and the density between the high- and low-velocity components, 
the spatial distributions of the temperature and the density are evident. 
At the high velocity,  
the gas temperature around the central protostar is a factor of two lower than that outside (Fig. \ref{tnmap} (a)), 
and the density is high at the center and decreases outward (Fig. \ref{tnmap} (b)). 
At the low velocity, 
the overall distributions of the temperature and the density are similar to those at the high velocity although their absolute values are lower and higher than those at the high velocity, respectively. 
In addition, 
the lower-temperature and higher-density regions at the low velocity are elongated along the north-south direction.

In addition to the temperature and the density,
the optical depths and the excitation temperatures of the CS (7--6) and C$^{18}$O (2--1) emission were also estimated.
In the high-velocity range, 
both the CS and C$^{18}$O lines are optically thin ($\tau_{CS}$ $<$0.1 \& $\tau_{C^{18}O}$ $<$0.2).
In the low-velocity range, 
both the CS and C$^{18}$O lines are optically thick ($\tau_{CS}$ $\sim$1.1 \& $\tau_{C^{18}O}$ $\sim$4.1) at the center and optically thin ($\tau_{CS}$ \& $\tau_{C^{18}O}$ $<$0.7) at a radius larger than $\sim$400 AU.
In both velocity ranges, 
the CS emission is sub-thermalized and has an excitation temperature of $\sim$10 to 20 K, 
while the C$^{18}$O emission is thermalized.

In the present LVG analysis, 
we assumed the constant C$^{18}$O and CS abundances in the entire region. 
In reality, 
those molecular abundances likely vary over the observed region (e.g., Tafalla et al. 2002, Takakuwa et al. 2000, 2011). 
To check the effect of the abundance variations on the estimate of the physical conditions, 
additional LVG analysis was performed with the molecular abundance as a free parameter, 
but a constant gas temperature of 20 K, 
which is the averaged temperature derived from the above-mentioned LVG analysis. 
It was found that the CS abundance needs to be comparable to the C$^{18}$O abundance in the high-velocity component, 
to reproduce the observed CS intensities higher than the observed C$^{18}$O intensities. 
This is highly unlikely from astrochemical point of view (e.g., Bergin and Langer 1997, Aikawa et al. 2008), 
and in fact previous multi-molecular line observations of B335 have revealed that the CS abundance is more than one order of magnitude lower than the C$^{18}$O abundance \citep{Eva05}. 
We therefore consider that the observed CS intensities higher than the C$^{18}$O intensities in the high-velocity components are mostly due to higher gas temperature of the high-velocity components compared with the low-velocity components regardless possible molecular abundance variations between the two components. 
Note that the relative difference in the gas temperature and density shown in Figure \ref{lvg} and \ref{tnmap} is still valid independently of the assumed molecular abundances.
The absolute values of the temperature and density for each component,  
however, still depend on the assumed abundances. 
Typically a factor of two variation of the abundances yields $\sim$a factor of two change of the gas density and $\sim$50\% change of the gas temperature in the low-velocity components, and a factor of two change of the gas density and temperature in the high-velocity component.

\section{Discussion}

\subsection{Physical Conditions of the Outflow and the Envelope in B335}
Our LVG analysis show that there are two types of gas components with different physical conditions in B335: 
the high-velocity ($|V|$ $\gtrsim$0.3 km s$^{-1}$), higher-temperature ($>$40 K), and lower-density ($<$10$^{6}$ cm$^{-3}$) component with the stronger submillimeter CS emission, 
and the lower-velocity ($|V|$ $\lesssim$0.3 km s$^{-1}$), lower-temperature ($<$40 K), and higher-density ($>$10$^{6}$ cm$^{-3}$) component with the stronger millimeter C$^{18}$O emission. 
As shown in Figure \ref{tnmap}, 
the stronger CS emission in the high-velocity component shows an elongated structure along the east-west direction, 
or along the direction of the molecular outflow associated with B335. 
This morphology as well as its higher velocity suggests that the stronger CS emission at the high velocity is related to the outflow. 
In contrast, 
the low-velocity component with the stronger C$^{18}$O emission is extended over the entire region with a compact structure surrounding the protostar, 
and shows a lower temperature and a higher density. 
Because of these characteristics, 
the low-velocity gas with the stronger C$^{18}$O emission is naturally considered to be originated from the protostellar envelope. 
Note that the compact structure at the low velocity shows an elongated structure perpendicular to the outflow. 
This fact also supports our suggestion that the low-velocity gas with the stronger C$^{18}$O emission is originated from the protostellar envelope.

In the envelope, 
the estimated gas density increases from 10$^{6}$ to 10$^{7}$ cm$^{-3}$ as the radius decreases from $\sim$1500 to $\sim$400 AU, 
and the estimated temperature decreases from 40 to 10 K.
\cite{Mor95} have made a single-dish survey for a complete flux-limited IRAS sample of protostellar sources in Taurus in the CS (3--2, 5--4, \& 7--6) and H$_{2}$CO (3$_{03}$--2$_{02}$ \& 3$_{22}$-2$_{21}$) emission lines with the beam size of $\sim$20$\arcsec$ to $\sim$40$\arcsec$ (1400 to 2800 AU in radius),
and have performed LVG analyses. 
They have found that the typical temperature and density of the sample protostellar envelopes are 20 to 50 K and a few $\times$ 10$^{6}$ cm$^{-3}$, respectively, 
approximately consistent with our high-resolution estimates of the physical conditions of the envelope around B335.
In the high-velocity component, 
the estimated gas density is $\lesssim$10$^{6}$ cm$^{-3}$, 
and the estimated gas temperature is higher than 40 K and even higher than 90 K in some parts.
The critical density of the submillimeter CS (7--6) emission is $\sim$2 $\times$ 10$^{7}$ cm$^{-3}$,
one order of magnitude higher than the estimated density in the high-velocity component.
In fact, 
the excitation temperature of the CS (7--6) emission in the high-velocity component is $\lesssim$20 K, 
more than a factor of two lower than the estimated gas kinetic temperature, 
and hence the submillimeter CS emission is sub-thermalized. 
On the other hand, 
the upper-state rotational energy of the submillimeter CS (7--6) emission is 66 K, 
comparable to the estimated gas temperature in the outflow component. 
These results suggest that the presence of the submillimeter CS emission on a scale larger than 1000 AU
toward B335 is not due to the presence of the extended high-density gas but due to the presence of the extended high-temperature gas.
Previous single-dish surveys for low-mass protostellar sources have found intense ($>$1 K) submillimeter molecular lines such as the CS (7--6), HCN (4--3) \citep{Mor95,Tak07, Tak11} and CO (7--6 \& 6--5) lines \citep{Sch93, Hog98, van06, van09, Shi09}.
The origin of the submillimeter molecular lines with extent of the single-dish beam sizes (1000 to 3000 AU) was unclear,
and mainly three mechanisms were proposed to explain the presence of the extended submillimeter molecular-line emission toward low-mass protostellar sources;
(1) shock heating associated with outflows,
(2) heating by ultraviolet photons from the boundary layer between a protostar and its surrounding disk, 
and (3) heating by protostellar luminosities \citep{Sch93, Spa95, van09}.
The latter two proposed mechanisms predict a higher temperature toward the innermost part of the envelope, 
inconsistent with our higher-resolution observational results in B335 showing a lower temperature in the innermost envelope.
Thus, we suggest that the origin of the submillimeter molecular-line emission on a scale larger than 1000 AU toward low-mass protostellar sources is related to the associated outflows, 
which presumably interact with the surrounding envelopes.

\subsection{Rotation, Infall, and Evolution of Protostellar Envelopes\\
Traced by the Millimeter and Submillimeter Line Emissions}

Our single-dish and SMA observations of B335 have revealed kinematics of the envelope from large ($\sim$10000 AU) to small ($\sim$400 AU) scales uniformly. 
The SMT C$^{18}$O (2--1) image shows a north (redshifted) -- south (blueshifted) velocity gradient across the outflow axis, 
which is consistent with the C$^{18}$O (1--0) result \citep{Sai99} and can be interpreted as a rotational motion in the outer envelope ($>$2500 AU). 
In the inner part, 
the SMA C$^{18}$O (2--1) and CS (7--6) images show the $\sim$1500 and $\sim$900 AU scale compact components associated with the protostar, respectively. 
The compact C$^{18}$O (2--1) component exhibits a velocity gradient along the outflow axis but no signature of a velocity gradient perpendicular to the outflow axis. 
The velocity gradient seen in the compact C$^{18}$O component along the outflow axis has been interpreted as an infalling motion, 
while there is no clear sign of rotation in the compact C$^{18}$O component (Paper I). 
In contrast, 
the compact CS component does not show a clear sign of the infalling motion, 
but show a hint of a velocity gradient across the outflow axis (i.e., rotation), 
although it is quite marginal. 
The rotational velocity and specific angular momentum of the compact CS component were estimated to be 0.2 km s$^{-1}$ and 8 $\times$ 10$^{-5}$ km s$^{-1}$ pc at a radius of 90 AU, respectively, 
from the mean offset of the peak positions of the compact CS component with respect to the protostellar position ($\sim$90 AU) and the north-south velocity gradient ($\sim$2 $\times$ 10$^{-3}$ km s$^{-1}$ AU$^{-1}$) seen in the CS velocity channel maps.
In this final subsection, 
we will discuss the rotation, infall, and the evolution of the protostellar envelope around B335.

In Figure \ref{jvsr}, 
we plot the specific angular momenta ($j$ = $r$ $\times$ $v_{\rm rot}$) of the envelope rotation as a function of the radius in B335, 
measured from our observations and from \cite{Sai99} (red diamonds in Figure \ref{jvsr}),  
and $r$ was derived from the radii of the emission sizes except the data point of the CS (7--6) emission, i.e., the data point at $r$ = 90 AU. 
For the data point of the CS (7--6) emission, 
$r$ was derived from the mean offset of the peak positions with respect to the protostellar position in the velocity channel maps. 
Since there is no clear velocity gradient in the central compact C$^{18}$O component across the outflow axis, 
the data point at the radius of 370 AU provides only the upper limit of the specific angular momentum of the envelope rotation (Paper I). 
Similarly, the CS data point at the radius of 90 AU may be an upper limit  
since the north-south velocity gradient in the CS emission is also marginal.  
Thus, 
these data points show that the specific angular momentum in the envelope around B335 decreases down to the radius of 370 AU. 
If the norht-south velocity gradient in the compact CS component is real and represents the rotation of the innermost envelope, 
the specific angular momentum at a radius smaller than 370 AU may be conserved.

For comparison, 
we also plot the specific angular momenta of other Class 0 (green squares), I (light blue squares) and II (brown squares) sources and NH$_{3}$ cores (gray squares; Goodman et al. 1993). 
For the data points of the NH$_{3}$ cores in Figure \ref{jvsr}, 
$r$ is the radii of the cores. 
For other sources, 
if Keplerian disks were observed in those sources, 
$r$ represents the outer radii of the Keplerian disks; 
if not, 
$r$ was derived from the radii of the emission sizes. 
The data points of the NH$_{3}$ cores show that the specific angular momenta at the outer radius ($r$ $>$10000 AU) decrease, 
and appear to follow the power-law relation of $j$ $\propto$ $r^{-1.6}$ \citep{Goo93}. 
On the other hand, 
specific angular momenta of the envelopes around Class 0, I, and II sources, including B335, spread over two orders of magnitude between $r$ $\sim$80 and $\sim$2000 AU. 
Furthermore, 
more evolved sources tend to show higher specific angular momenta than less evolved sources (i.e., from Class 0 to II), 
as already reported in Paper I.

\cite{Oha97b} have proposed that the distribution of the specific angular momenta as a function of the radius in low-mass dense cores and protostellar envelopes consists of two zones; 
one with the power-law distribution ($j$ $\propto$ $r^{-1.6}$) at $r$ $>$6000 AU, 
and the other with the common, conserved angular momentum ($j$ $\sim$10$^{-3}$ km s$^{-1}$ pc) associated with the infalling motion at $r$ $<$6000 AU. 
Our results of B335, and the data plots of more Class 0, I, and II sources show that the angular momenta of Class 0, I, and II sources within $<$2000 AU are not common but dependent on the evolutionary stage of the central sources.

One possible scenario of the evolution of the specific angular momentum profile is as follows. 
We consider the case that the initial distribution of the specific angular momentum in a dense core follows the power-law relation as observed by \cite{Goo93}.  
Once the inside-out collapse takes place in such a dense core, 
the innermost part of the envelope with the smaller specific angular momentum accretes to the center. 
Hence, 
in the earlier stage of the collapse, 
the infalling material has a smaller specific angular momentum. 
As the expansion wave propagates outward, 
material in the outer region of the envelope with the larger angular momentum starts collapsing. 
Therefore, 
the more evolved sources tend to have higher angular momenta and larger regions with a constant specific angular momentum. 

In order to demonstrate the above scenario quantitatively, 
the expected evolution of the distribution of the specific angular momenta in protostellar envelopes (black dashed curves in Figure \ref{jvsr}) was calculated,
based on the inside-out collapse of an isothermal spherical envelope \citep{Shu77} with the initial profile of the specific angular momentum ($j$) of   
\begin{equation}
j = 6.1 \times 10^{22} \cdot r_{\rm rot}^{1.6}\ {\rm cm^{2}\ s^{-1}}, 
\end{equation} 
where $r_{\rm rot}$ is the rotational radius \citep{Goo93}. 
Here, 
we consider the time evolution of the envelope material at the initial radius of $r_{\rm int}$, 
whose specific angular momentum is expressed by equation (2).
The enclosed mass within $r_{\rm int}$ can be calculated by integrating the following initial density profile of the spherical envelope, 
\begin{equation}
\rho = \frac{C_{\rm s}}{2\pi G}r^{-2},
\end{equation}
where $C_{\rm s}$ is the sound speed and $G$ is the gravitational constant \citep{Shu77}. 
The sound speed was adopted to be 0.2 km s$^{-1}$ as in the case of isothermal envelopes at $\sim$10 K. 
Note that in reality protostellar envelopes are not isothermal as we discussed in section 4. 
Since our main interest is to estimate how much the specific angular momentum is transfered by the infalling material, 
we focus on the material that is infalling through the equator plane where the largest specific angular momentum is transfered. 
In order to obtain the radial profile of the specific angular momentum at a given time, 
it is required to calculate the position of the infalling material on the equator plane.
For this purpose, 
we simply assumed that each infalling material follows free-fall motion, 
and that there is no effect of centrifugal force and pressure gradient. 
On this assumption, 
the motion of the infalling material on the equator plane can be described as, 
\begin{equation}
\ddot{r} = \frac{GM}{r^{2}}, 
\end{equation}
where $M$ is the enclosed mass within $r_{\rm int}$, 
calculated from equation (3).
Then we can calculate the radial position of each infalling material with each specific angular momentum as a function of time.
For the material that is located at a radius of 1000 AU on the equator plane in the initial envelope, 
when the material starts to fall in,
its gravitational energy is more than one order magnitude larger than the rotational energy unless the material reaches a radius of 10 AU,  
so we did not calculate the distribution of the specific angular momentum within a radius of 10 AU in the envelope.  
The infalling motion is terminated artificially once the material passes through $r$ = 10 AU.
In the calculation, 
the material was set to fall after the passage of the expansion wave, 
and we stopped the calculation at $t$ = 3 $\times$ 10$^{5}$ years after the beginning of the collapse.  
The expansion wave reaches at a radius of $\sim$13000 AU at that time.

Four black dashed curves in Figure \ref{jvsr} show the calculated profiles of the specific angular momentum on the equator plane of the envelope at $t$ = 2, 4, 7, 30 $\times$ 10$^{4}$ years after the beginning of the collapse. 
All the profiles show similar curves, 
except for the radius where the specific angular momentum starts shifting from the initial specific angular momentum distribution. 
This radius corresponds to the radius of the expansion wave. 
For example, at 2 $\times$ 10$^{4}$ years, 
the expansion wave reaches a radius of $\sim$800 AU, 
and the material at that radius starts infalling. 
Once the material starts infalling, 
its angular momentum is carried toward the center, 
so the profile of the specific angular momentum starts shifting from the initial specific angular momentum distribution, 
and shows a constant specific angular momentum in the infalling envelope.   
With time evolution the material at an outer radius with a larger specific angular momentum starts infalling, 
and the specific angular momentum of the infalling envelope increases. 
Eventually
in the innermost region the rotational motion starts to dominate over the infalling motion, 
and a Keplerian disk is expected to form. 
The sizes of Keplerian disks could grow from less than ten to hundreds of AU 
when protostellar sources evolve from Class 0 to II stages, 
as discussed in Paper I.
The profile of the constant specific angular momentum in an infalling envelope eventually evolves to the Keplerian profile. 
The transition from the infalling motion with the conserved angular momentum to the Keplerian rotation is, however, still not well understood.

\section{Summary}
We have performed detailed imaging and analyses of the combined SMA + single-dish data of B335 in the millimeter C$^{18}$O (2--1) and submillimeter CS (7--6) emission lines. Main results obtained from our millimeter and submillimeter imaging of the low-mass protostellar envelope around B335 are summarized below. 
\begin{enumerate}
\item{The C$^{18}$O emission traces both extended and central compact components with sizes of $\sim$9000 and $\sim$1500 AU, respectively,   
and the CS emission shows a central compact component with a size of $\sim$900 AU surrounded by an east-west elongated component with a size of $\sim$3000 AU. 
The extended C$^{18}$O component exhibits a north (redshifted) -- south (blueshifted) velocity gradient across the outflow axis, 
while the compact C$^{18}$O component exhibits an east (blueshifted) -- west (redshifted) velocity gradient along the outflow axis. 
The CS emission shows an east (redshifted) -- west (blueshifted) velocity gradient and also a possible north (redshifted) -- south (blueshifted) velocity gradient. 
At $|V|$ $\gtrsim$0.3 km s$^{-1}$ the CS emission is stronger and more extended than the C$^{18}$O emission, 
while the C$^{18}$O emission is stronger and more extended than the CS emission at $|V|$ $\lesssim$0.3 km s$^{-1}$. 
}
\item{With the millimeter C$^{18}$O (2--1) and submillimeter CS (7--6) lines,
we estimated physical conditions of molecular gas in B335 based on the LVG model. 
At the high velocity ($|V|$ $>$0.3 km s$^{-1}$), 
the gas component with the stronger CS emission shows a temperature higher than 40 K, a density lower than 10$^{6}$ cm$^{-3}$, and elongation along the east-west direction, or along the outflow axis.
This high-temperature and low-density component is likely related to the outflow, 
and the presence of the submillimeter CS (7--6) emission on a scale larger than 1000 AU is likely due to an extended high-temperature region but not an extended high-density region.  
At the low velocity ($|V|$ $<$0.3 km s$^{-1}$), 
the gas component with the stronger C$^{18}$O emission shows a temperature lower than 40 K and a density higher than 10$^{6}$ cm$^{-3}$, 
and is elongated perpendicular to the outflow axis. 
This low-temperature and high-density component is likely originated from the envelope.  
In the entire region, 
the C$^{18}$O (2--1) emission is thermalized ($T_{\rm ex} \sim T_{\rm k}$), and the CS (7--6) emission is sub-thermalized ($T_{\rm ex} \sim$10--20 K $<$ $T_{\rm k}$). 
On the other hand, 
the C$^{18}$O (2--1) and CS (7--6) lines are optically thin at the high velocity, 
while both lines are optically thick at the center and optically thin at a radius larger than $\sim$400 AU at the low velocity. 
Regardless of the adopted value of the molecular abundances, 
the higher temperature in the high-velocity component with the more intense submillimeter CS emission than that in the low-velocity component is probably valid, 
since otherwise we have to adopt the unusually high CS abundance comparable to the C$^{18}$O abundance in the high-velocity component. 
The absolute values of the temperature and density for each component,  
however, still depend on the assumed abundances.
}
\item{The velocity gradient across the outflow axis seen in the extended C$^{18}$O emission component can be interpreted as a large-scale ($>$2500 AU) envelope rotation. 
On the other hand, 
in the compact C$^{18}$O component no signature of the envelope rotation is seen, 
but there exists a velocity gradient along the outflow axis, 
which can be interpreted as an infalling gas motion. 
The compact CS component does not show a clear sign of the infalling motion, 
but show a possible sign of a velocity gradient across the outflow axis (i.e., rotation). 
From these observational results, 
we constructed the profile of the specific angular momentum of the envelope rotation in B335, from radii of $\sim$10$^{4}$ AU down to 10$^{2}$ AU, and found that the specific angular momentum decreases down to a radius of 370 AU.
}
\item{Specific angular momenta of envelopes around a sample of Class 0, I, and II sources, including B335, spread over two orders of magnitude between $r$ $\sim$80 and $\sim$2000 AU, 
and more evolved sources tend to show higher specific angular momenta than less evolved sources. 
If the initial profile of the specific angular momentum in a dense core follows the power-law relation as observed by \cite{Goo93} and the inside-out collapse takes place in the core, 
the inner infalling region with less specific angular momenta should collapse first and show the region with the small and constant specific angular momentum. 
At a later stage of the inside-out collapse, 
the outer region with more specific angular momenta starts collapsing and forming the larger infalling region with the larger specific angular momenta. 
From these considerations we constructed a toy model to explain the observed profiles of the specific angular momenta and their evolutions.
}
\end{enumerate}

\acknowledgments
We would like to thank all the SMT, ASTE, and SMA staff supporting this work. The research of S. T. and N. O. are supported by NSC 99-2112-M-001-013-MY3 and NSC99-2112-M-001-008-MY3, respectively. 

\appendix
\section{Combining SMA and Single-dish Data and \\
Imaging Simulation of the Combining Process}
To combine the SMA and single-dish data, 
we followed the process described by \cite{Tak03,Tak07}, 
which is based on the description of combining single-dish and interferometric data by \cite{Vog84} 
and the MIRIAD scripts developed by \cite{Wil94}. 
First, we resampled the single-dish image cube along the velocity axis to match the velocity channels with those of the SMA data. 
Then we de-convolved the single-dish images by the single-dish beams and multiplied the de-convolved single-dish images by the response function of the SMA primary beams, 
which we assumed to be two-dimensional Gaussians.
With these de-convolved and primary beam de-corrected images, 
we generated single-dish visibility data by the Miriad tasks, $uvrandom$ and $uvmodel$. 
Next, the single-dish and SMA visibility data are Fourier-transformed simultaneously by the Miriad task, $invert$, to make the combined images. 
The SMA C$^{18}$O and CS visibility data have 13440 data points per channel at $u$--$v$ distances from 5 to 50 $k\lambda$ and 9985 data points per channel at $u$--$v$ distances from 12 to 80 $k\lambda$, 
respectively, 
and we varied the numbers of the visibility data points of our single-dish data to make the combined images and adopted 4866 data points per channel at $u$--$v$ distances from 0 to 7 $k\lambda$ and 1963 data points per channel at $u$--$v$ distances from 0 to 9 $k\lambda$ for the SMT and ASTE data, respectively. 
Our weighting on the single-dish data is two to five times more than the weighting recommended by \cite{Kur09}. 
Although the recommended weighting can provide smaller beam sizes and higher signal-to-noise ratios, 
the recommended weighting introduced lots of patchy structures in the velocity channel maps, 
which could be due to the influence of the noise in the SMA data. 
Therefore, 
we adopted a higher weighting on our single-dish data than the recommended value to smooth out those patchy structures.

In order to test the feasibility and the limitation of the combining process described above, 
we performed noise-free imaging simulations and compared an original model image with its final image after the combining process. 
We generated an model image of a single power-law intensity distribution, $I(r)$ $\propto$ $r^{-1}$, 
which represents the intensity distribution of an isothermal and optically thin sphere with a density profile of $r^{-2}$.
We set the outer radius of the intensity distribution to be 15000 AU (100$\arcsec$ at a distance of 150 pc) and did not include any velocity structure in the model. 
The image pixel size and the pixel number of the model image were set to be 0\farcs2 $\times$ 0\farcs2 and 2048 $\times$ 2048, respectively.
We adopted two virtual single-dish telescopes with the dish sizes of 10 and 30 meters  
and the observing frequency of the C$^{18}$O (2--1) line.
The beam sizes of the virtual 10-m and 30-m telescopes are 33\farcs8 and 11\farcs3, respectively.
The virtual 10-m single-dish telescope corresponds to SMT,
and the simulation with the virtual 10-m telescope represents our real imaging.
First, we convolved this model image with the single-dish beams and resampled the convolved model images onto the Nyquist grids (15$\arcsec$ for the 10-m telescope and 5$\arcsec$ for the 30-m telescope) over our SMT mapping region that is 13500 AU $\times$ 18000 AU. 
After we produced the simulated single-dish images, 
we followed the same process described in the last paragraph to create the simulated single-dish visibility data.
Next, we multiplied the original model image by the response function of the SMA primary beam 
and generated the simulated SMA visibility data with the $u$--$v$ sampling of our SMA observation at 230 GHz.
In Figure \ref{uvamp}, we plot the amplitudes of our simulated visibility data as a function of the $u$--$v$ distance. 
The amplitude of the simulated 10-m single-dish data is below that of the simulated SMA data in the $u$--$v$ distance range of 5--6 k$\lambda$.
On the other hand, 
the amplitude of the simulated 30-m single-dish data well matches that of the simulated SMA data in the $u$--$v$ distance range of 5--8 k$\lambda$, 
but becomes lower than that of the simulated SMA data at the longer $u$--$v$ distances. 

To derive the correct solution of the amplitude as a function of the $u$--$v$ distance for the original model image, 
we applied SMA primary-beam de-correction to the original model image and Fourier-transformed the entire de-corrected image over the $u$--$v$ sampling from 0 to 20 k$\lambda$ (solid curves in Figure \ref{uvamp} left). 
In the right panel of Figure \ref{uvamp}, 
we show the amplitude ratio between the simulated single-dish data and the correct solution as a function of the $u$--$v$ distance.
The amplitudes of the simulated 10-m and 30-m single-dish data are 10\% and 5\% lower than that of the correct solution at the zero spacing, respectively, 
and a factor of five lower than that of the correct solution at their maximum $u$-$v$ sampling distances.
This mismatch between the single-dish amplitude and the correct solution at the zero spacing is likely due to the limited mapping region of the single-dish observations, which does not cover the entire model intensity distribution.
On the other hand, the difference at the longer $u$--$v$ distances is likely due to the finite single-dish beam sizes and the grid spacings 
since the sharpness of the image is diluted. 
Furthermore, 
the amount of the mismatch between the correct solution and the simulated amplitudes in the $u$--$v$ domain should be structure-dependent.
Therefore, 
a simple scaling of the amplitude cannot correct the discrepancy of the amplitude between single-dish and interferometric data properly, 
but artificially distorts the total flux. 

Figure \ref{simmap} compares the simulated 10-m single-dish + SMA image after the combining process to the original model image multiplied by the SMA primary beam and convolved with the same synthesized beam of the simulated combined image. 
The combined image exhibits a central compact structure plus an extended component, 
different from the original single power-law intensity distribution. 
Figure \ref{simprofile} compares the intensity profiles as a function of the radius along the north-south and east-west directions. 
The intensity of the central region at a radius less than 10$\arcsec$ becomes weaker after the combining process, 
while that in the outer region is amplified. 
As a result, the original single power-law intensity distribution appears to be separated into two components.
The flux in the inner region ($r$ $<$5\arcsec) of the combined image is suppressed by $\lesssim$20\% of the peak flux of the combined image, 
and the difference between the original and combined images at a radius larger than 5\arcsec is less than 10\% of the peak flux of the combined image.
The amount of the intensity suppression in the central region within 5$\arcsec$ ($\sim$20\%) is, however, comparable to our uncertainty of the absolute flux calibrations of the single-dish and SMA observations, 
and the maximum difference between the "real" and "observed" images at a radius larger than 5$\arcsec$ is likely to be 10\% of the peak flux at most as shown in the present simulation, 
which corresponds to 1.5$\sigma$ in our real observations.

\section{LVG Model}
For our LVG calculations, 
values of the dipole moment and the rotational and centrifugal constants of C$^{18}$O and CS were taken from \cite{Win79}.
The collisional coefficients of C$^{18}$O were taken from \cite{Flo01}, 
on the assumption that C$^{18}$O has the same collisional coefficients as those of $^{12}$CO.
The rotational energy levels up to $J$ = 28 (2087 K) were included in our C$^{18}$O calculations.
The collisional coefficients of CS were taken from \cite{Gre78} and \cite{Tur92},
and the rotational energy levels up to $J$ = 20 (447 K) were included.
Since we only have one transition of each of C$^{18}$O and CS,
the physical conditions and the molecular abundances cannot be derived simultaneously.
Instead, 
we estimated typical values of the molecular abundances per unit velocity gradient ($\tbond$ $\frac{X}{dV/dR}$)  
and adopted those $\frac{X}{dV/dR}$ values to estimate the gas density and temperature in the different velocity components.
The values of $\frac{X}{dV/dR}$ were estimated as follows.
The full width at the half maximum line width of the ASTE CS line profile at the center was measured to be $\sim$1.5 km s$^{-1}$ \citep{Tak07}.
In the combined moment 0 map of the CS emission, the size of the CS emitting region was estimated to be $\sim$3000 AU (0.015 pc).
Therefore, 
$dV$ and $dR$ were set to be 1.5 km s$^{-1}$ and 0.015 pc, respectively,
and the velocity gradient was estimated to be 100.0 km s$^{-1}$ pc$^{-1}$. 
Since the LVG model a priori assumes that the C$^{18}$O and CS emission originate from the same region, 
the same velocity gradient was adopted for both lines. 

For the C$^{18}$O abundance we adopted $X$(C$^{18}$O) = 2.5 $\times$ 10$^{-8}$, from Table 6 in \cite{Eva05}.
This C$^{18}$O abundance is one order of magnitude lower than the typical C$^{18}$O abundance that was estimated with a single-dish beam size of $\sim$1$\farcm$6 (3 $\times$ 10$^{-7}$; Frerking et al. 1987),
and a similar C$^{18}$O abundance has also been found with our SMA observation (3.8 $\times$ 10$^{-8}$; Paper I).
For the CS abundance
we first adopted $X$(CS) = 6 $\times$ 10$^{-9}$, again from Table 6 in \cite{Eva05}.
Then, the $\frac{X}{dV/dR}$ values are 2.5 $\times$ 10$^{-10}$ km$^{-1}$ s pc and 6 $\times$ 10$^{-11}$ km$^{-1}$ s pc for C$^{18}$O and CS, respectively.
We found, 
however, 
that the observed peak brightness temperature of the C$^{18}$O emission ($\sim$6.1 K) and that of the CS emission ($\sim$4.3 K) toward the center cannot be reproduced simultaneously with these $\frac{X}{dV/dR}$ values.
Furthermore, 
our LVG model provided the estimate of the gas kinetic temperature lower than 10 K in the region at radii from $\sim$400 to $\sim$1500 AU with these $\frac{X}{dV/dR}$ values, which is unphysical, 
since the previous millimeter continuum study of the B335 envelope shows the temperature from 22 to 13 K at radii from $\sim$400 to $\sim$1500 AU \citep{Har03}.
On the other hand, 
previous interferometric observations of B335 in the CS (5--4) emission have found that the CS abundance could be lower than the typical value ($\sim$6 $\times$ 10$^{-9}$) by a factor of $\sim$ten \citep{Wil00}.
\cite{Eva05} have also suggested a possible one order of magnitude lower CS abundance in B335,
based on the modeling of the single-dish line profiles of the CS (2--1, 3--2, \& 5--4) emission lines with the density and temperature profiles found by \cite{Har03}. 
In the case of L1551 IRS5, a Class I protostar, 
\cite{Tak04} has found that the CS abundance is also lower than the typical value by a factor of ten.
Therefore, we adopted the CS abundance of 6 $\times$ 10$^{-10}$ in our LVG analyses,
and the adopted $\frac{X}{dV/dR}$ values are 2.5 $\times$ 10$^{-10}$ km$^{-1}$ s pc and 6 $\times$ 10$^{-12}$ km$^{-1}$ s pc for C$^{18}$O and CS, respectively. 
If the typical C$^{18}$O and CS abundances (i.e., $X$({C$^{18}$O) = 3.0 $\times$ 10$^{-7}$ and $X$(CS) = 6.0 $\times$ 10$^{-9}$) were adopted, 
the estimated temperature is a factor of $\sim$three higher, 
and the estimated density is a factor of $\sim$ten lower.

\begin{figure}
\epsscale{1}
\plotone{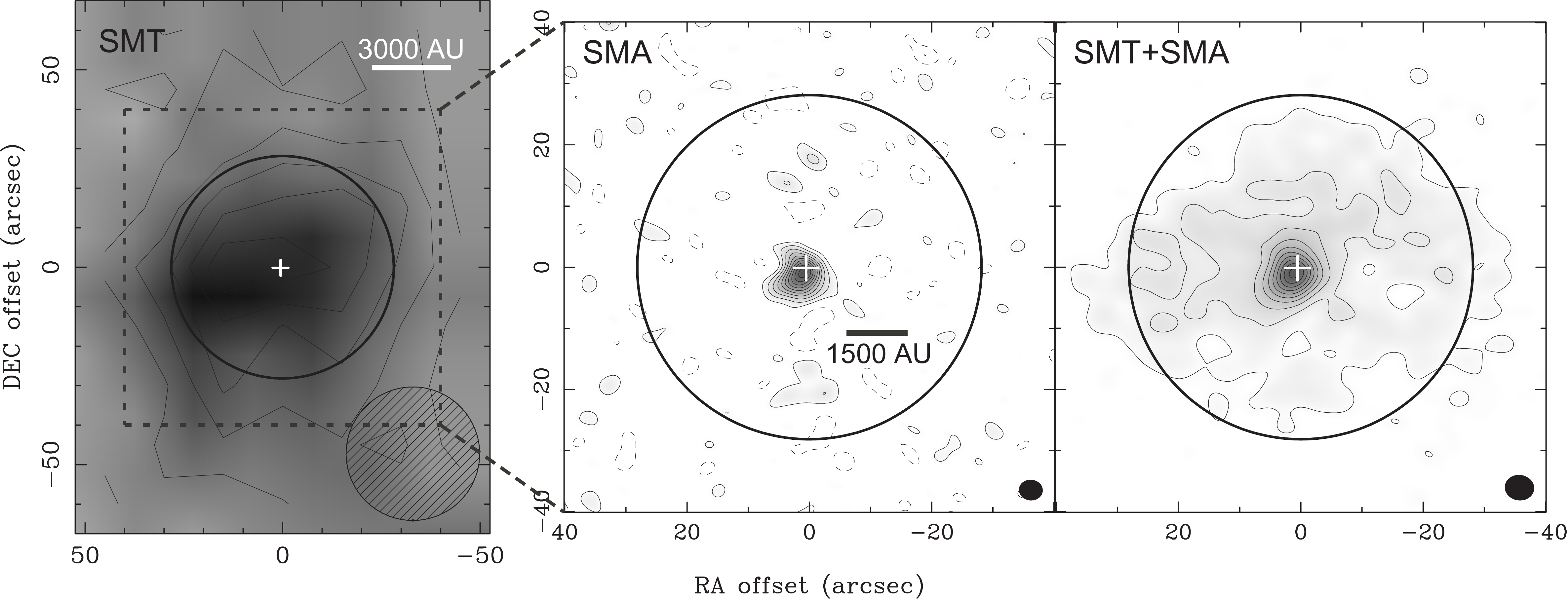}
\caption{Moment 0 maps of the SMT (left), SMA (middle), and combined (right) data of the C$^{18}$O (2--1) emission in B335. 
The integrated velocity ranges are $V_{\rm LSR}$ = 7.95 -- 8.92, 7.29 -- 9.23, and 7.57 -- 8.95 km s$^{-1}$ for the SMT, SMA, and combined images, respectively. 
Crosses show the protostellar position, and open circles represent the primary beam of the SMA observation at 230 GHz. 
Hatched and solid ellipses at the bottom-right corners show the beam sizes. Contour levels are from 14$\sigma$ in steps of 3$\sigma$ in the SMT map, where 1$\sigma$ is 0.05 K km s$^{-1}$. The SMA and combined maps are plotted in the same contour levels and gray scale, and contour levels are from 0.85 K km s$^{-1}$ in steps of 0.85 K km s$^{-1}$ that corresponds to 2$\sigma$ in the SMA map and 2.2$\sigma$ in the combined map. The peak values are 1.57, 8.91, and 8.73 K km s$^{-1}$ in the SMT, SMA and combined maps, respectively.}\label{c18omom0}
\end{figure}

\begin{figure}
\epsscale{1}
\plotone{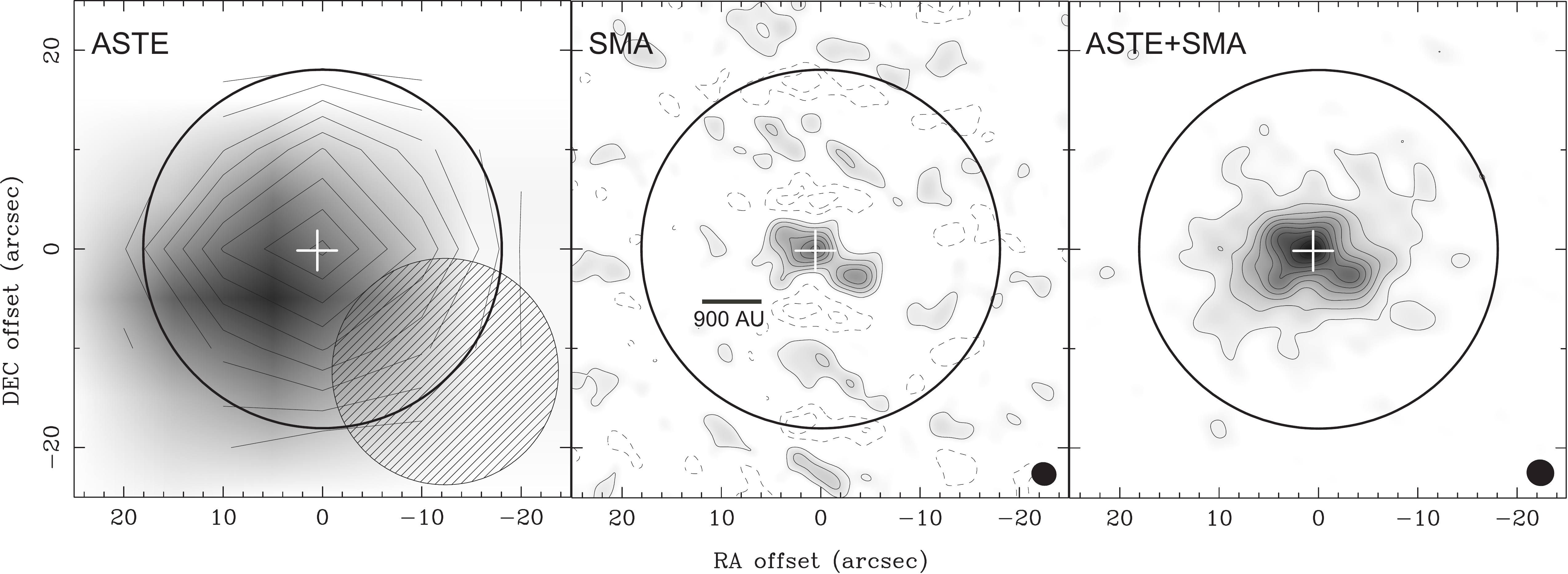}
\caption{Moment 0 maps of the ASTE (left), SMA (middle), and combined (right) data of the CS (7--6) emission in B335. 
The integrated velocity ranges are $V_{\rm LSR}$ = 7.18 -- 9.05, 7.47 -- 9.25, and 7.12 -- 9.25 km s$^{-1}$ for the ASTE, SMA, and combined images, respectively.
Crosses show the protostellar position, and open circles represent the primary beam of the SMA observation at 342 GHz. Hatched and solid ellipses at the bottom-right corners show the beam sizes. Contour levels are from 3$\sigma$ in steps of 3$\sigma$ in the ASTE map, where 1$\sigma$ is 0.06 K km s$^{-1}$. The SMA and combined maps are plotted in the same contour levels and gray scale, and contour levels are from 1.02 K km s$^{-1}$ in steps of 1.02 K km s$^{-1}$ that corresponds to 2.4$\sigma$ in the SMA map and 2$\sigma$ in the combined map. The peak values are 1.67, 6.39, and 9.91 K km s$^{-1}$ in the ASTE, SMA and combined maps, respectively.}\label{csmom0}
\end{figure}

\begin{figure}
\epsscale{1}
\plotone{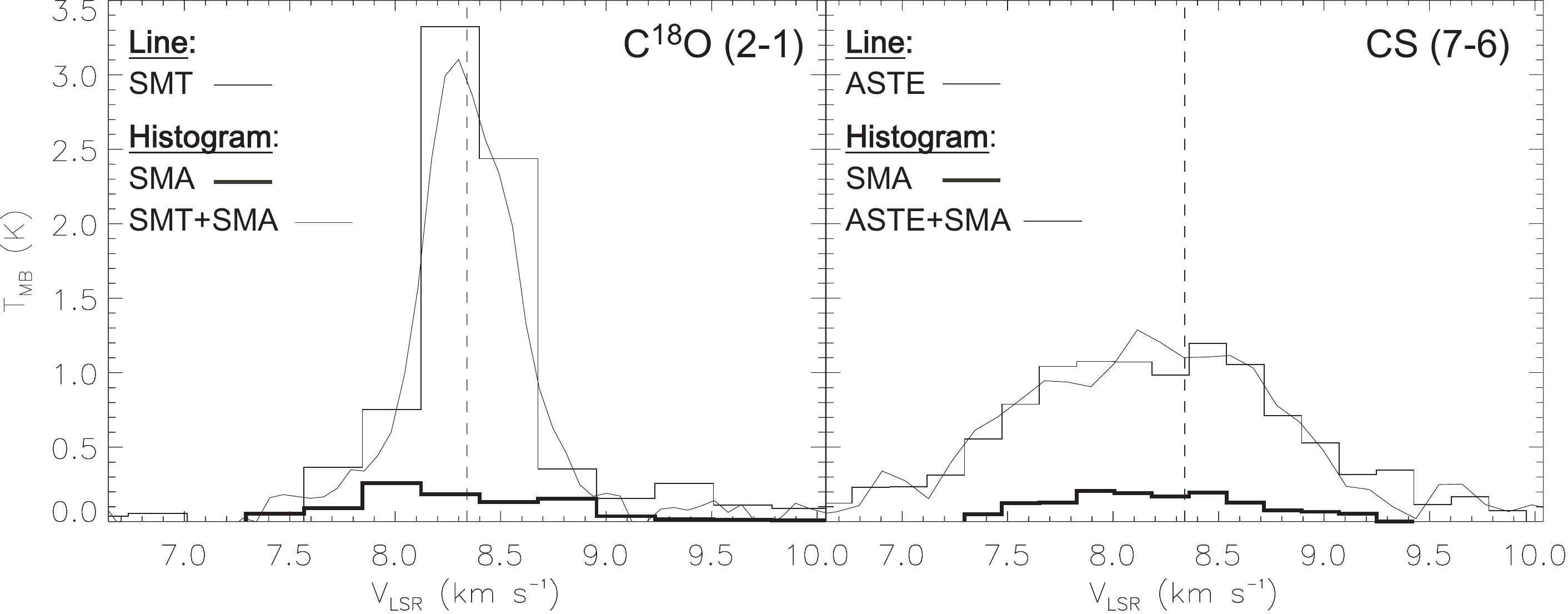}
\caption{Line profiles of the C$^{18}$O (2--1) emission in the SMT, SMA, and combined data (left) and the CS (7--6) emission in the ASTE, SMA, and combined data (right) toward the protostellar position. The single-dish line profiles are drawn in thin lines, and the line profiles of the SMA and combined data are plotted in thick and thin histograms, respectively. In order to compare the line profiles between the single-dish, SMA, and combined data, the SMA and combined data were first primary-beam corrected and then convolved with the relevant single-dish beam sizes. The beam sizes of the SMT and ASTE observations are 33\farcs8 and 21\farcs6, respectively. Vertical dashed lines show the systemic velocity of B335, and the systemic velocity was estimated to be $V_{\rm LSR}$ = 8.34 km s$^{-1}$ from a Gaussian fitting to the SMT C$^{18}$O (2--1) emission.}\label{line}
\end{figure}

\begin{figure}
\epsscale{0.7}
\plotone{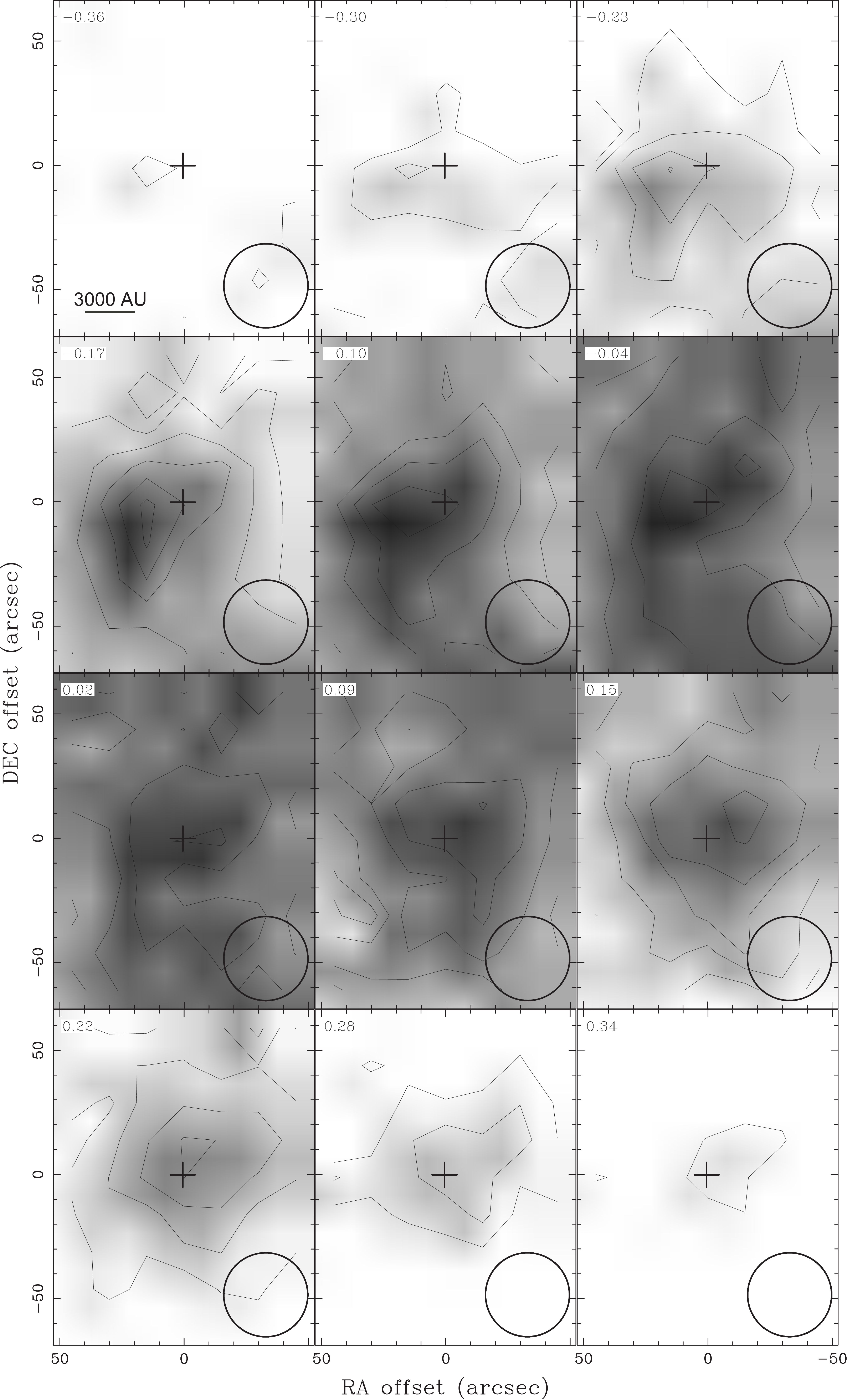}
\caption{Velocity channel maps of the SMT C$^{18}$O (2--1) data in B335. The central velocity at each channel is shown at the top-left corner of each panel. Crosses and open circles show the protostellar position and the beam size, respectively. Contour levels are from 3$\sigma$ in steps of 2$\sigma$, where 1$\sigma$ is 0.22 K.}\label{smtchannel}
\end{figure}

\begin{figure}
\epsscale{1}
\plotone{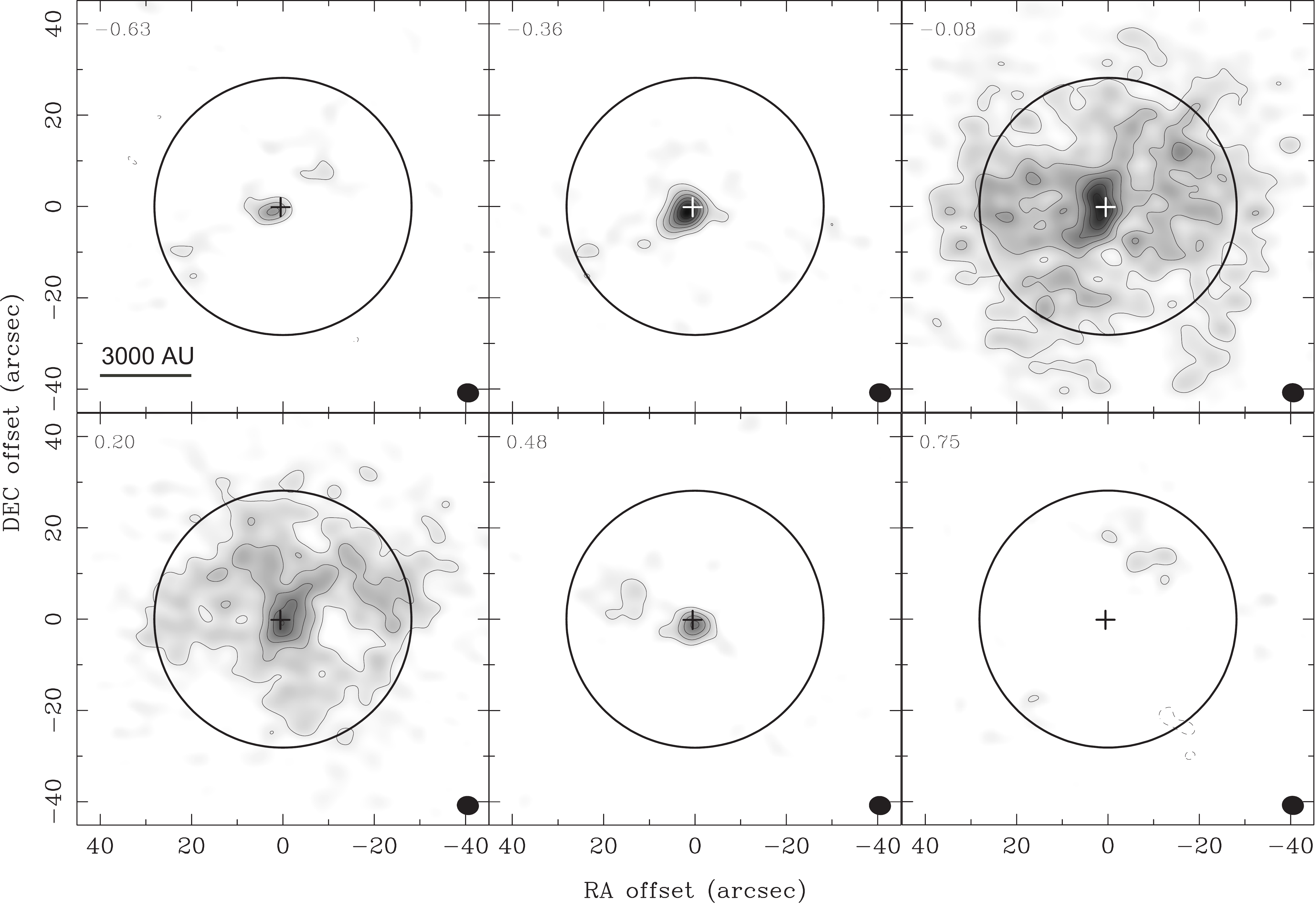}
\caption{Velocity channel maps of the combined C$^{18}$O (2--1) data in B335. The central velocity at each channel is shown at the top-left corner of each panel. Crosses and filled ellipses show the protostellar position and the beam size, respectively, and open circles are the primary beam of the SMA observation at 230 GHz. Contour levels are from 2$\sigma$ in steps of 2$\sigma$, where 1$\sigma$ is 0.63 K.}\label{c18ochannel}
\end{figure}

\begin{figure}
\epsscale{1}
\plotone{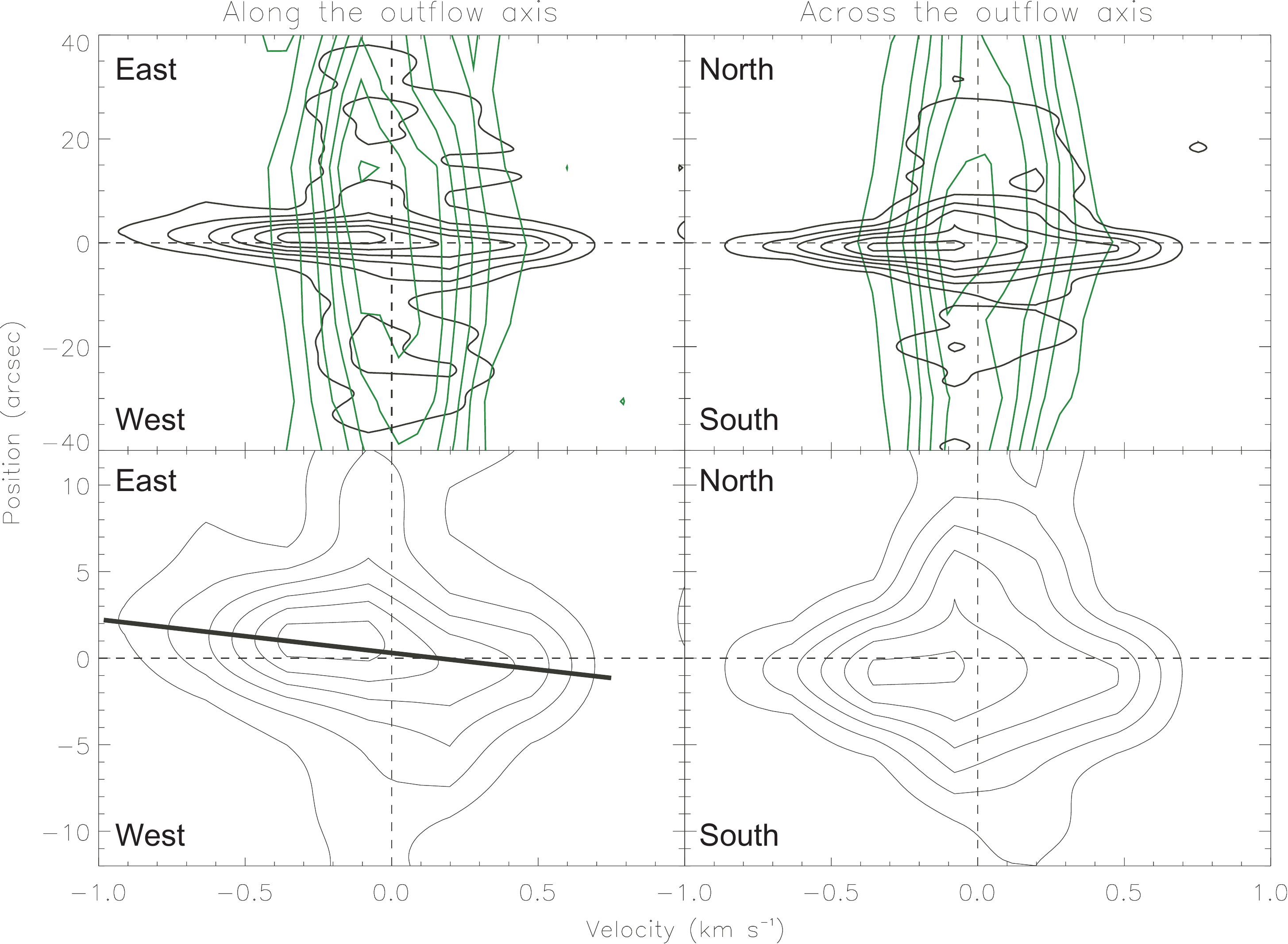}
\caption{SMT and combined SMT + SMA $P$--$V$ diagrams of the C$^{18}$O (2--1) emission along (P. A. = 90\degr; left) and across (P. A. = 0\degr; right) the outflow axis in B335, passing through the protostellar position. Top panels show the $P$--$V$ diagrams of the SMT (green contours) and combined SMT + SMA (black contours) data on a larger scale ($\sim$12000 AU), and bottom panels show the $P$--$V$ diagrams of the combined SMT + SMA data on a smaller scale ($\sim$3000 AU). Vertical and horizontal dashed lines delineate the systemic velocity and the central protostellar position, respectively. A solid line in the bottom-left panel delineates the detected velocity gradient along the outflow axis. Contour levels are from 2$\sigma$ in steps of 2$\sigma$, where 1$\sigma$ are 0.22 and 0.63 K in the SMT and combined SMT + SMA data, respectively.}\label{c18opv}
\end{figure}

\begin{figure}
\epsscale{1}
\plotone{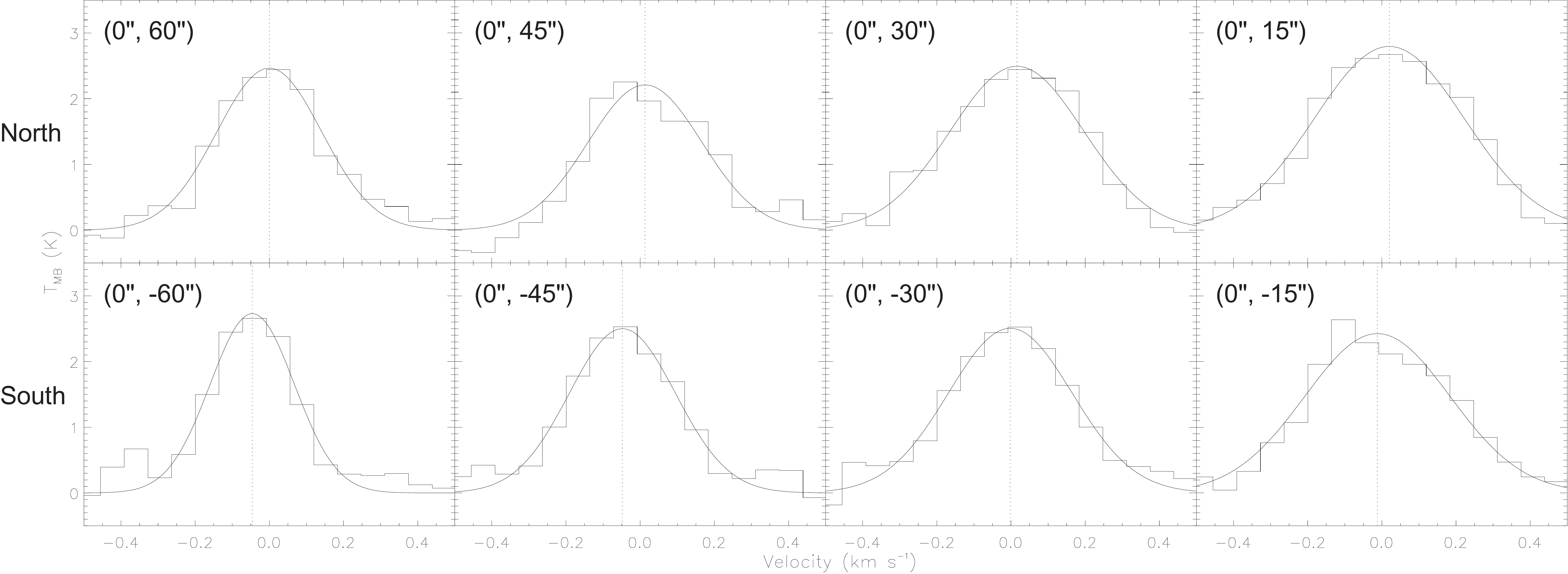}
\caption{SMT C$^{18}$O spectra (histogram) in B335 along the axis perpendicular to the outflow, i.e., the north-south direction. The offset position of each spectrum is shown at the top-left corner of each panel. Solid curves show the results of the Gaussian fitting to the SMT C$^{18}$O spectra, and a dotted line in each panel represents the centroid velocity of each spectrum derived from the Gaussian fitting.}\label{SMTspec}
\end{figure}

\begin{figure}
\epsscale{1}
\plotone{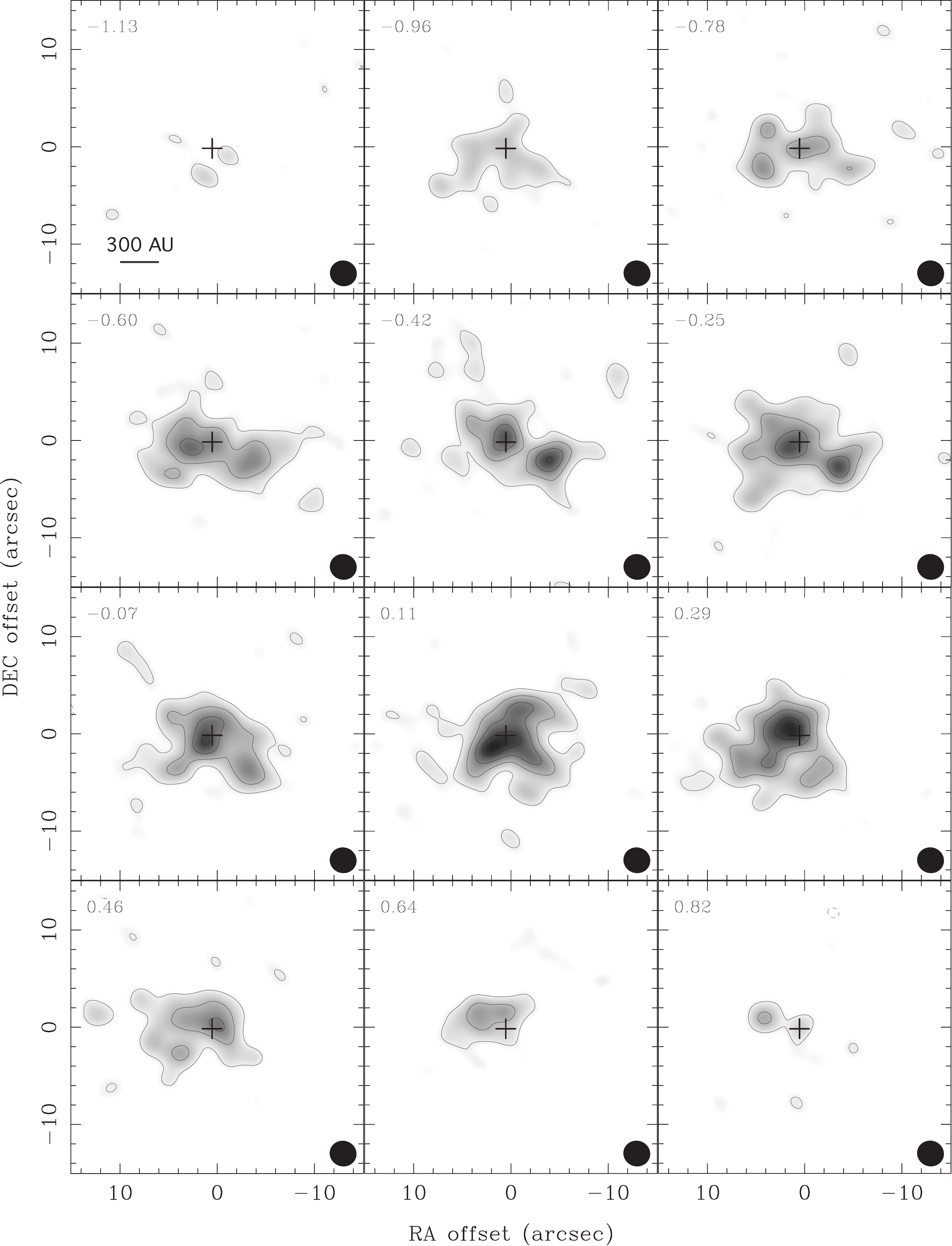}
\caption{Velocity channel maps of the combined CS (7--6) data in B335. The central velocity at each channel is shown at the top-left corner of each panel. Crosses and filled ellipses show the protostellar position and the beam size, respectively. Contour levels are from 2$\sigma$ in steps of 2$\sigma$, where 1$\sigma$ is 0.83 K.}\label{cschannel}
\end{figure}

\begin{figure}
\epsscale{1}
\plotone{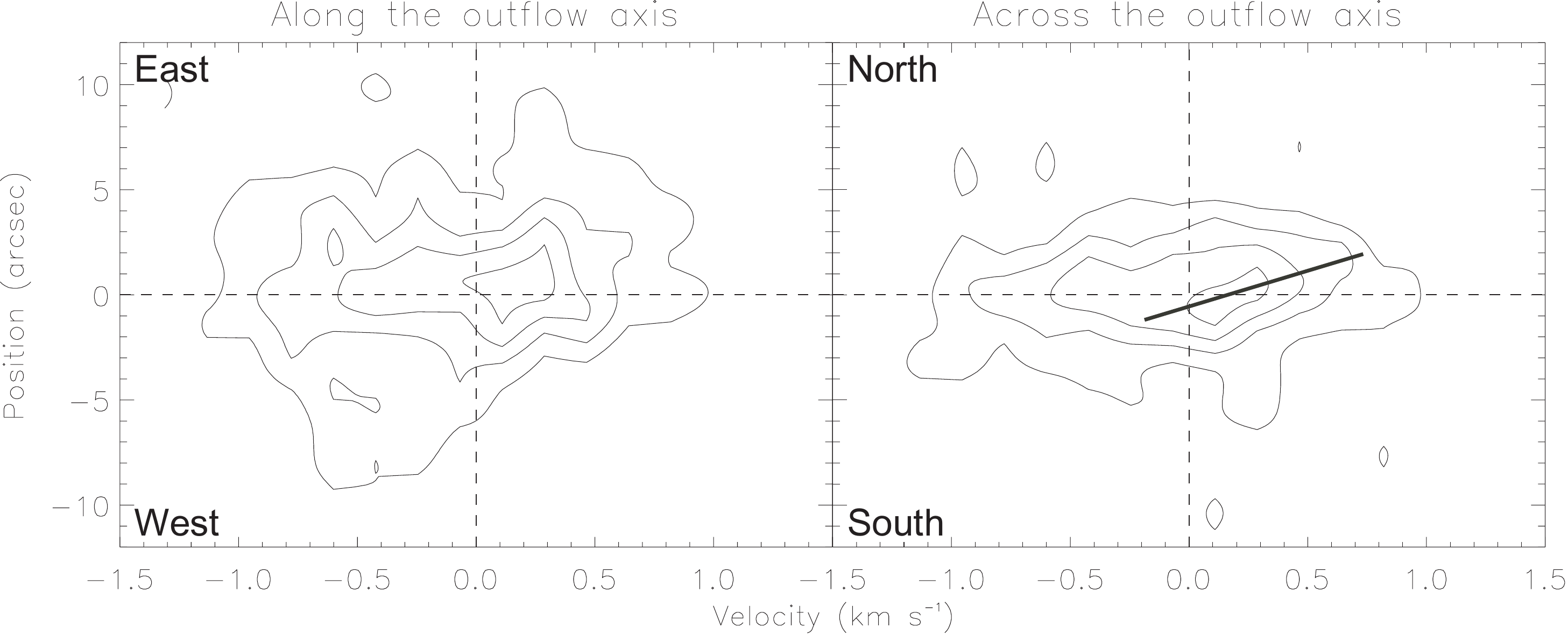}
\caption{combined ASTE + SMA $P$--$V$ diagrams of the CS (7--6) emission along (P. A. = 90\degr; left) and across (P. A. = 0\degr; right) the outflow axis in B335, passing through the protostellar position. Vertical and horizontal dashed lines delineate the systemic velocity and the central protostellar position, respectively.  A solid line in the right panel delineates the possible velocity gradient across the outflow axis seen in the velocity channel maps. Contour levels are from 2$\sigma$ in steps of 2$\sigma$, where 1$\sigma$ is 0.83 K.}\label{cspv}
\end{figure}

\begin{figure}
\epsscale{1}
\plotone{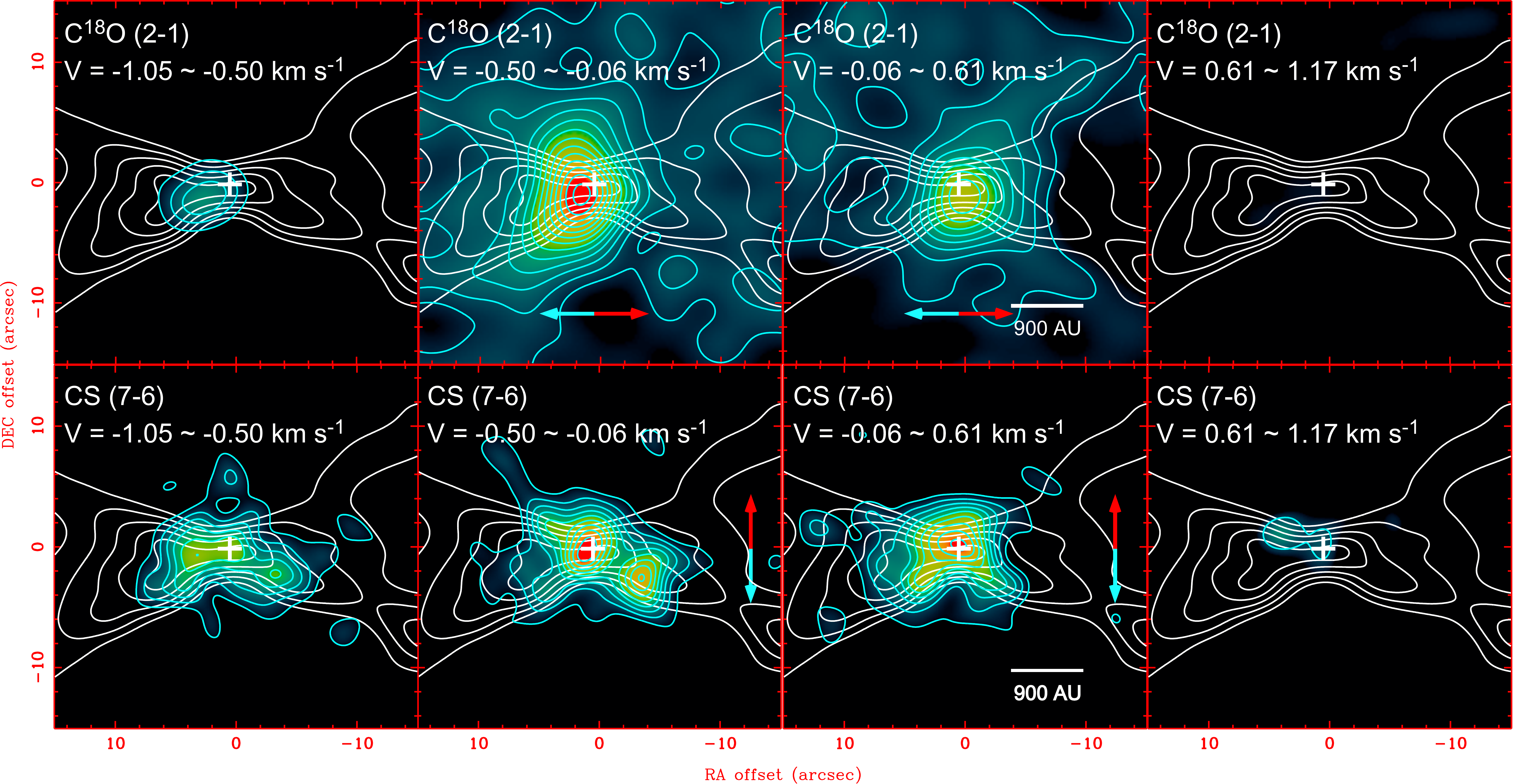}
\caption{Moment 0 maps (color scale and blue contours) of the C$^{18}$O (2--1) (top panels) and CS (7--6) (bottom panels) emission in B335 integrated over different velocity ranges as shown in the upper sides of the panels. White contours overlaid on these maps show the $^{12}$CO (2--1) moment 0 map taken with the SMA (Paper I). White crosses denote the protostellar position. Blue and red arrows represent the velocity gradients of the central components traced by the millimeter C$^{18}$O and submillimeter CS lines in the middle two panels. Contour levels are from 0.66 K in steps of 0.33 K, and 0.33 K corresponds to 1.3$\sigma$ in the top panels and 1$\sigma$ in the bottom panels.}\label{kinematics}
\end{figure}

\begin{figure}
\epsscale{1}
\plotone{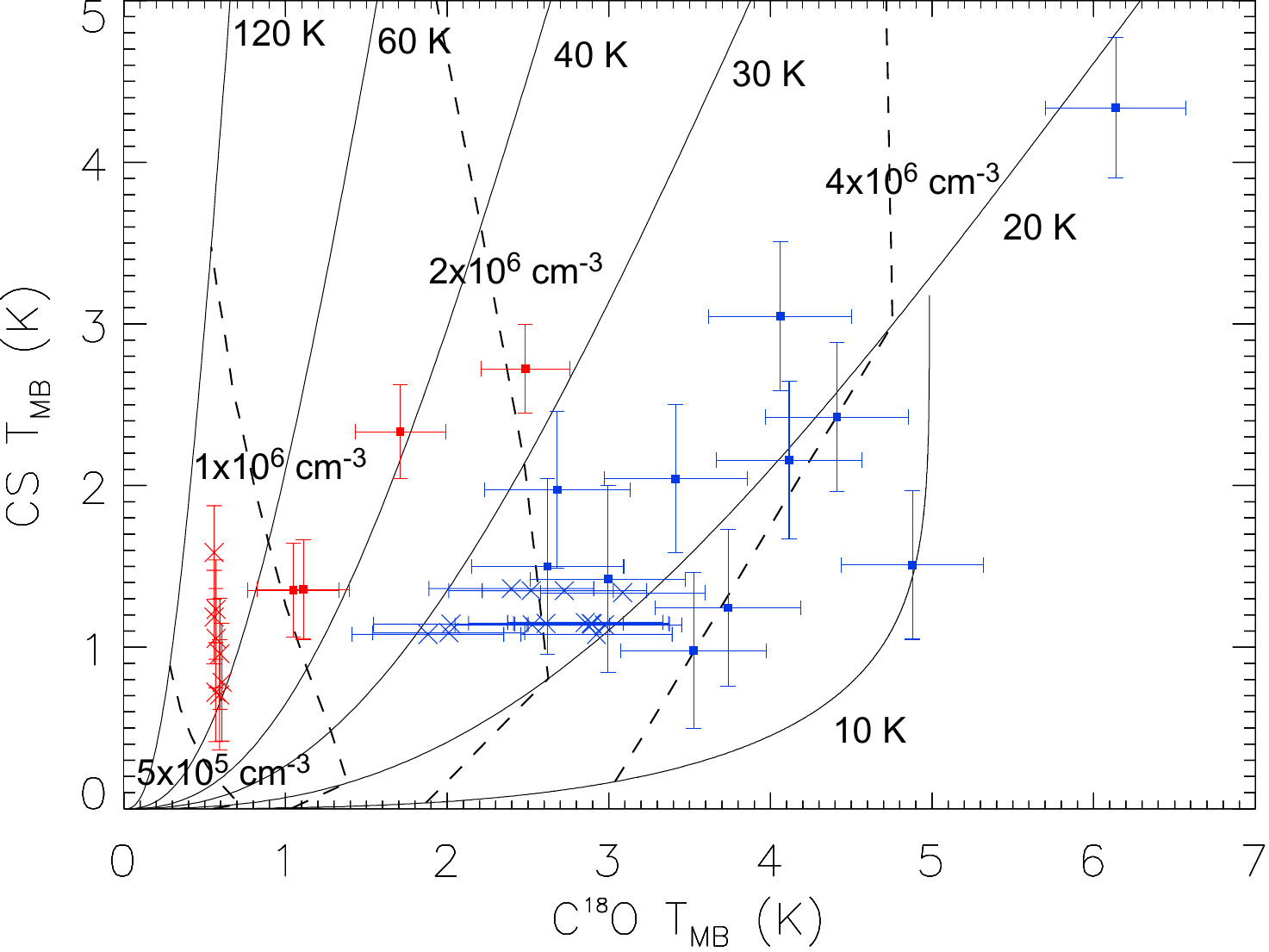}
\caption{Correlation diagram of the millimeter C$^{18}$O (2--1) and submillimeter CS (7--6) intensities in B335, overlaid on the iso-density (dashed curves) and iso-thermal (solid curves) contours calculated from our LVG model. All the data points were taken from the region within the SMA primary beam at 342 GHz. The integrated velocity ranges for the red points are from -1.05 to -0.22 and from 0.34 to 0.89 km s$^{-1}$, that is the high-velocity range. The integrated velocity range for the blue points is from -0.22 to 0.34 km s$^{-1}$, the low-velocity range. 
The data points labeled by crosses and without horizontal or vertical error bars represent the 2$\sigma$ upper limits of the C$^{18}$O or CS intensities, respectively. Error bars demote the 1$\sigma$ noise level.}\label{lvg}
\end{figure}

\begin{figure}
\epsscale{1}
\plotone{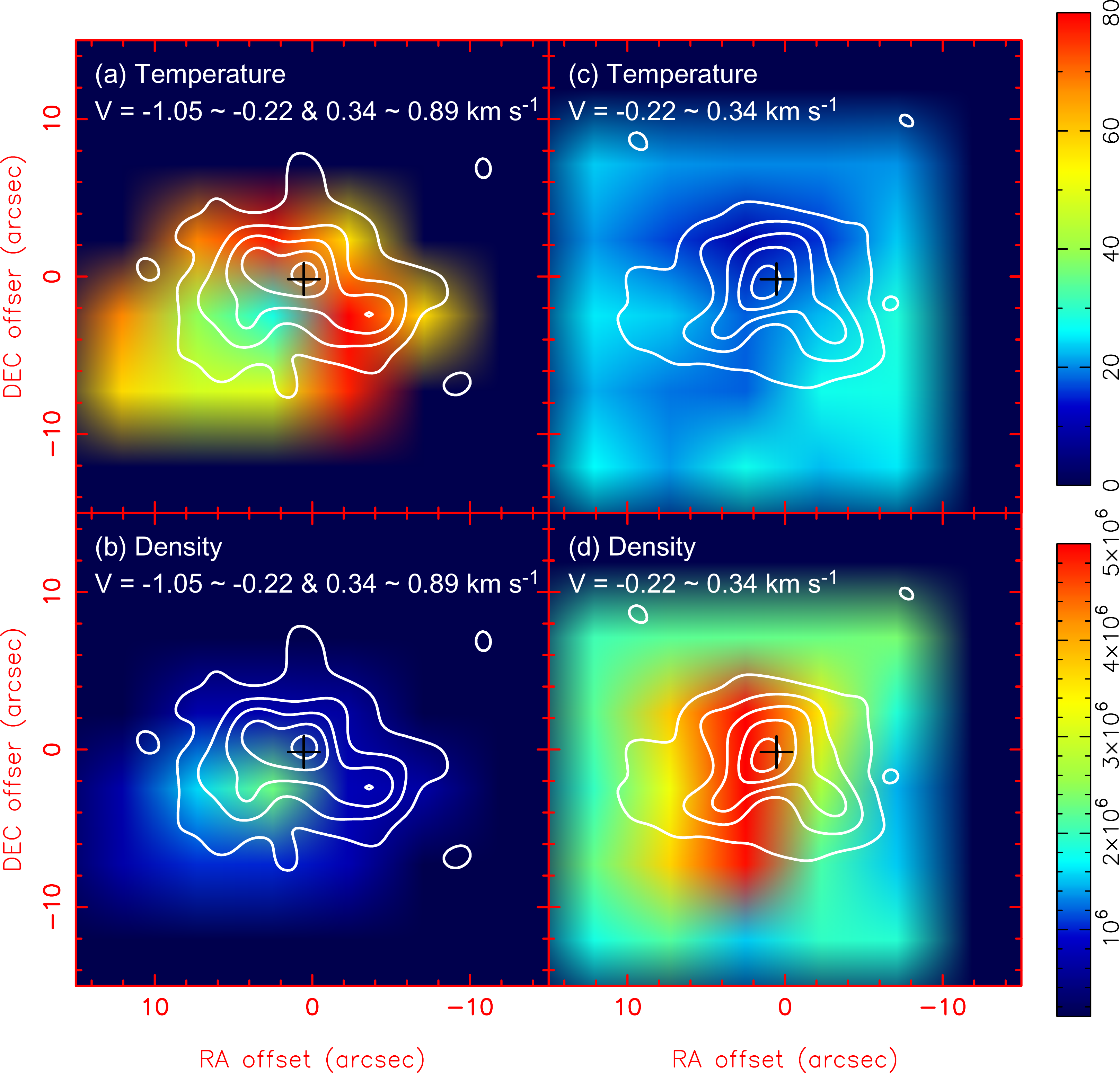}
\caption{Temperature (top) and density (bottom) maps at the high (left) and low (right) velocities in color scale, derived from our LVG calculations shown in Figure \ref{lvg}. White contours show the distribution of the CS emission integrated over the relevant velocity ranges. The integrated velocity range is shown in the upper side of each panel. Crosses represent the protostellar position. Contour levels are from 2$\sigma$ in steps of 2$\sigma$, where 1$\sigma$ is 0.38 K km s$^{-1}$ in the high-velocity range and 0.24 K km s$^{-1}$ in the low-velocity range.}\label{tnmap}
\end{figure}

\begin{figure}
\epsscale{0.8}
\plotone{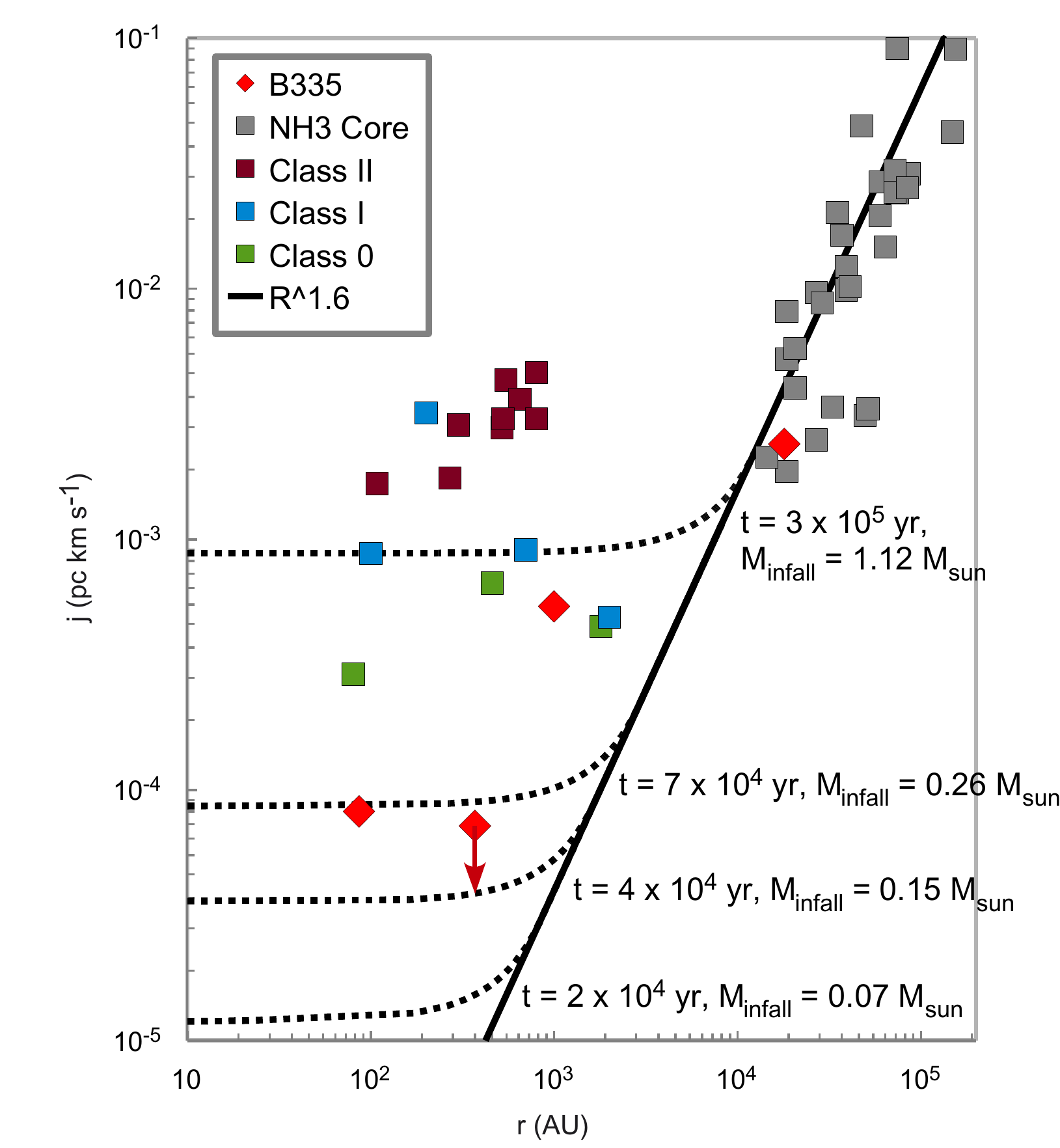}
\caption{Diagram of the specific angular momentum of the envelope rotation plotted as a function of the radius. Red diamonds show measurements of the envelope rotation in B335 (present paper; Paper I; Saito et al. 1999), and a red arrow shows the upper limit of the specific angular momentum of the envelope rotation at a radius of 370 AU in B335 (Paper I). Gray squares and a black solid line represent specific angular momenta in NH$_{3}$ cores and the power-law relation between the core sizes and the specific angular momenta, respectively \citep{Goo93}. Green, light blue, and dark red squares show specific angular momenta of disks or envelopes in Class 0, I and II sources, respectively \citep{Oha97a, Mom98, Sim00, Lee06, Che07, Lom08, Lee09}. Black dashed curves show the expected quantitative evolution of the profiles of specific angular momenta of infalling and rotating envelopes at t = 2, 4, 7, 30 $\times$ 10$^{4}$ years after the beginning of the collapse, and the masses within the infall radii are 0.07, 0.15, 0.26, and 1.12 M$_{\sun}$ at those times, respectively.}\label{jvsr}
\end{figure}

\begin{figure}
\epsscale{1}
\plotone{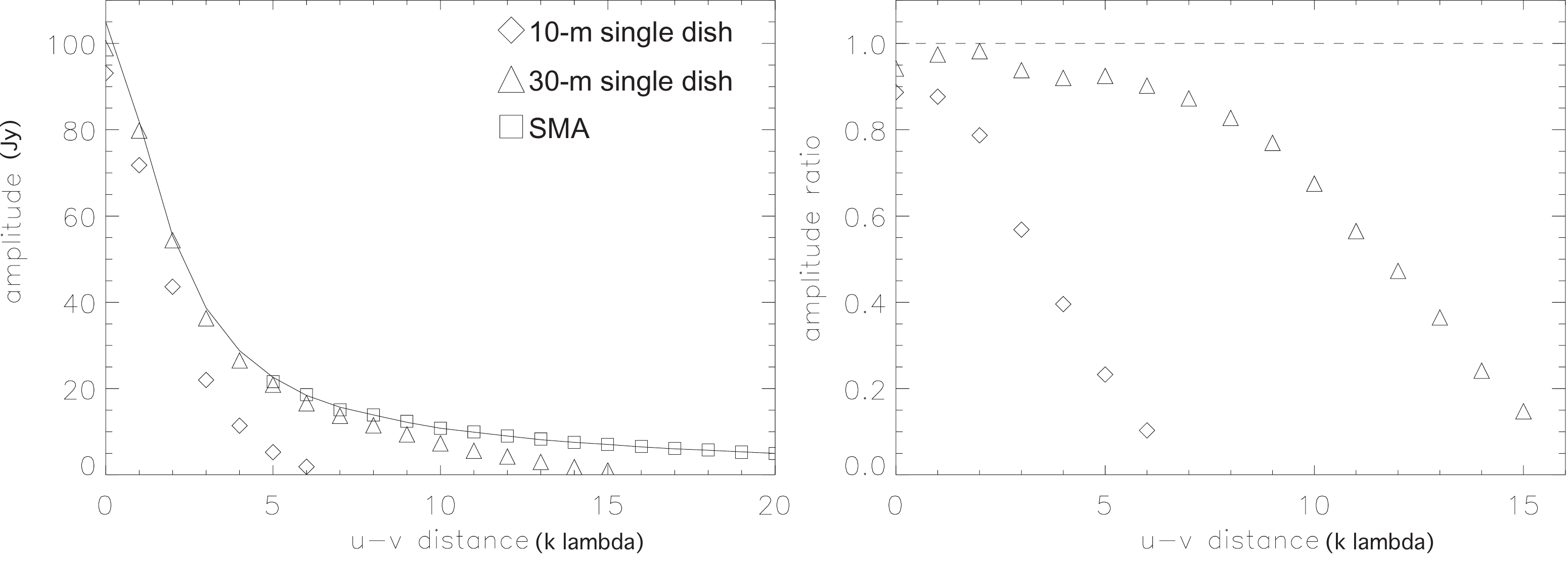}
\caption{Model and simulated amplitudes (left) and amplitude ratios (right) as a function of the $u$--$v$ distance, derived from our imaging simulations. Open diamonds, triangles, and squares show the simulated 10-m single-dish, 30-m single-dish, and SMA visibility data, respectively. A solid curve represents the amplitude of the model. The amplitude ratios were calculated by dividing the amplitude of the simulated single-dish data by that of the model.}\label{uvamp}
\end{figure}

\begin{figure}
\epsscale{1}
\plotone{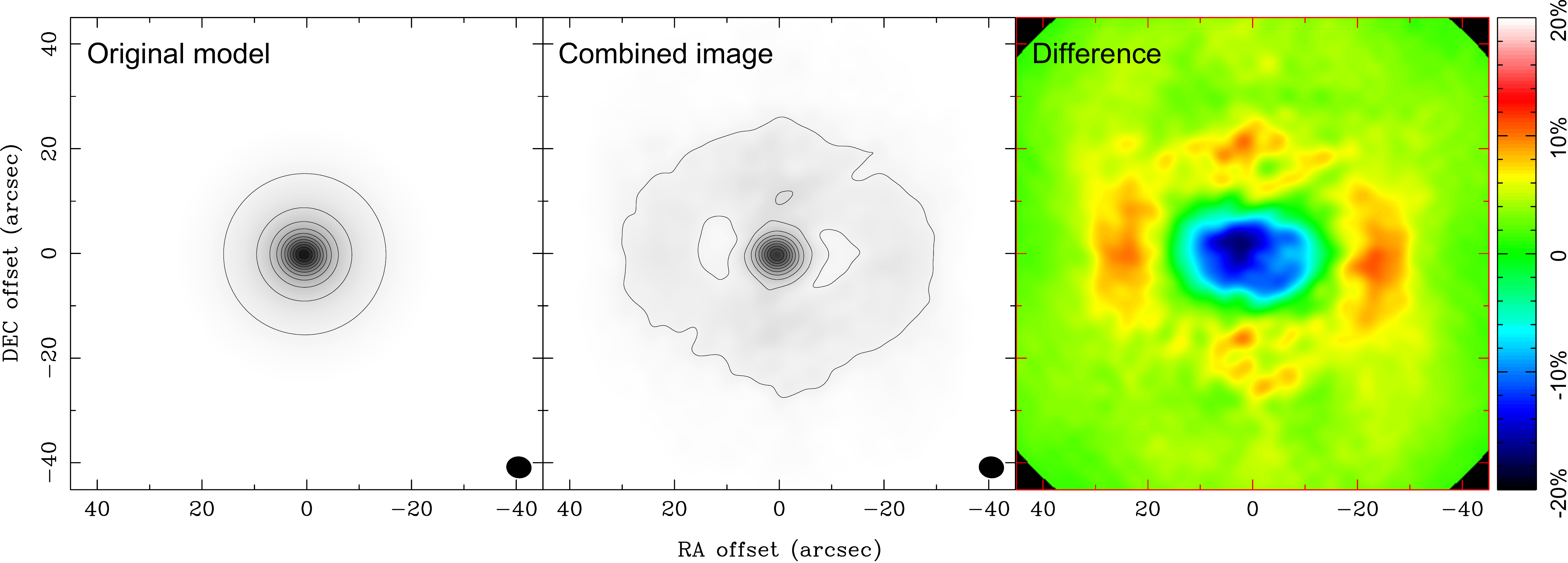}
\caption{Comparison of the model and simulated images from our imaging simulations. The left, middle, and right panels show the original model image, the simulated image after the combing process, and the difference between the model and simulated images, respectively. The combing process was performed with the 10-m single-dish and SMA data. The model image was de-corrected for the SMA primary beam and convolved with the simulated synthesized beam. Contour levels are from 10\% in steps of 10\% of the maximum intensity in the simulated combined image. A filled ellipse in each panel shows the beam size. The difference image is shown in the unit of percentages of the peak intensity of the simulated combined image.}\label{simmap}
\end{figure}

\begin{figure}
\epsscale{1}
\plotone{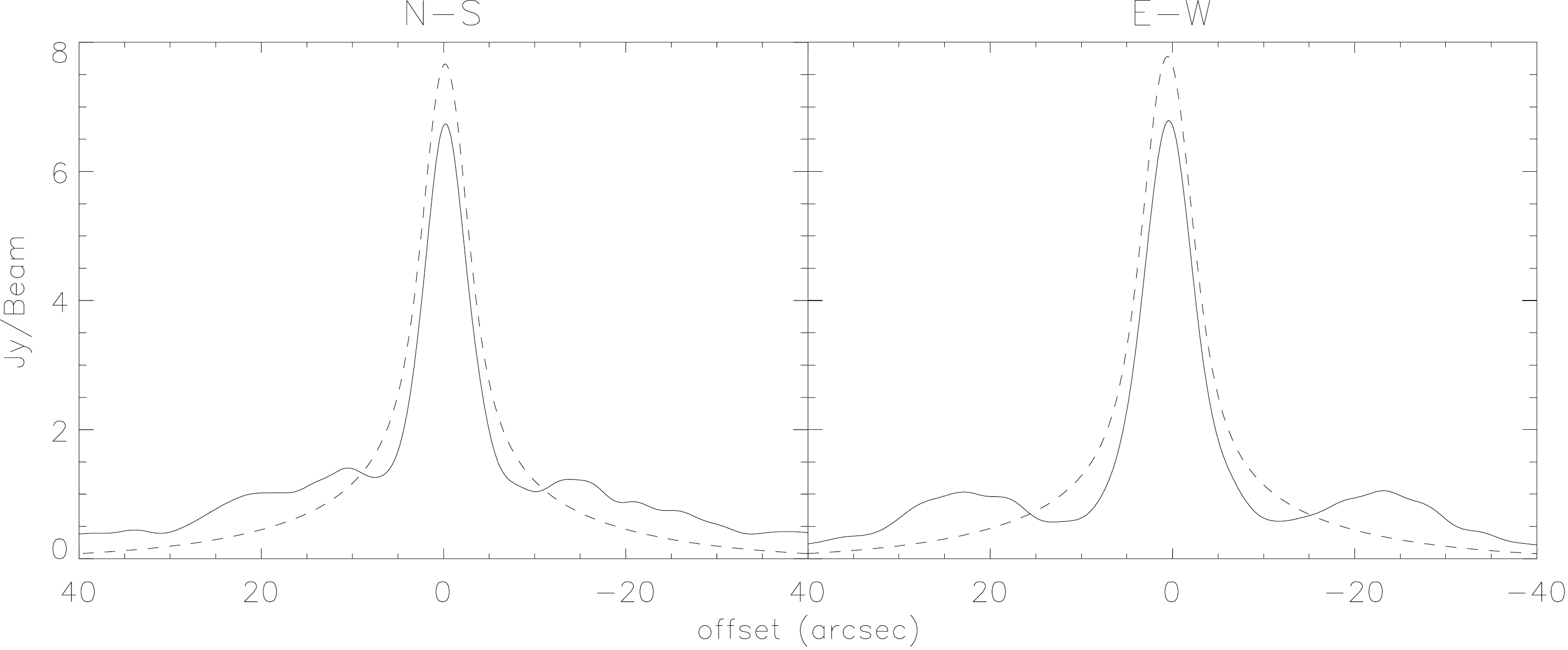}
\caption{Intensity profiles of the model image (dashed curves) and the simulated image after the combining process (solid lines) along the north-south (left) and east-west (right) directions. The combining process was performed with the 10-m single-dish and SMA data. The model image was de-corrected for the SMA primary beam and convolved with the simulated synthesized beam.}\label{simprofile}
\end{figure}

\clearpage

\begin{deluxetable}{lcc}
\tablewidth{0pt}
\tablecaption{Summary of the Observational Parameters \label{ob}}
\tablehead{ & C$^{18}$O (2--1) & CS (7--6)}
\startdata
Interferometer & \multicolumn{2}{c}{SMA} \\
\hline
Coordinate Center & \multicolumn{2}{c}{$\alpha$(J2000) = 19$^{h}$37$^{m}$00$\fs$89} \\
 & \multicolumn{2}{c}{$\delta$(J2000) = 7\arcdeg34\arcmin10$\farcs$0} \\
Primary Beam & 56\farcs3 & 36\farcs1 \\
Projected Baseline Length (k$\lambda$) & 5.5 -- 53.5 & 12.5 -- 80.0 \\ 
Synthesized Beam (P. A.) & 3\farcs8 $\times$ 3\farcs3 (84.6\degr) & 2\farcs5 $\times$  2\farcs3 (69.4\degr) \\
Velocity Resolution (km s$^{-1}$) & 0.28 & 0.18 \\
Noise Level (K) & 0.58 & 0.77 \\
Passband Calibrator & 3C 279 & 3C 454.3 \\
Flux Calibrator & Callisto & Uranus \\
Gain Calibrator (Flux) & 1749+096 (1.9 Jy) & 1749+096 (1.0 Jy) \\
 & 2145+067 (2.5 Jy) & 2145+067 (1.0 Jy) \\ \\
\hline
Single-dish Telescope & SMT & ASTE \\
\hline
Coordinate Center & \multicolumn{2}{c}{$\alpha$(J2000) = 19$^{h}$37$^{m}$00$\fs$90} \\
 & \multicolumn{2}{c}{$\delta$(J2000) = 7\arcdeg34\arcmin08$\farcs$8} \\
Beam Size & 33\farcs8 & 21\farcs6 \\
Velocity Resolution (km s$^{-1}$) & 0.06 & 0.11 \\
Main Beam Efficiency & 68\% & 60\% \\
Noise Level (K) & 0.22 & 0.14 \\ 
Conversion Factors (Jy Beam$^{-1}$ K$^{-1}$) & 66.3 & 74.8 \\ \\
\hline
Combined Data & SMT + SMA & ASTE + SMA \\
\hline
Coordinate Center & \multicolumn{2}{c}{$\alpha$(J2000) = 19$^{h}$37$^{m}$00$\fs$89} \\
 & \multicolumn{2}{c}{$\delta$(J2000) = 7\arcdeg34\arcmin10$\farcs$0} \\
Synthesized Beam (P. A.) & 4\farcs7 $\times$ 4\farcs0 (82.0\degr) & 2\farcs7 $\times$  2\farcs6 (72.3\degr) \\
Velocity Resolution (km s$^{-1}$) & 0.28 & 0.18 \\
Noise Level (K) & 0.63 & 0.83 
\enddata
\end{deluxetable}

\end{document}